\documentclass[%
reprint,
superscriptaddress,
showpacs,preprintnumbers,
 amsmath,amssymb,
 aps,
 prd,
]{revtex4-1}

\usepackage{graphicx}
\usepackage{adjustbox}
\usepackage{epstopdf}
\usepackage{epstopdf}
\usepackage{makebox}
\usepackage{lipsum}
\usepackage{mathtools}
\usepackage{subfig,float}
\usepackage[skip=5pt,font={bf,sc,small},singlelinecheck=off]{caption}
\begin{document}


\title{Quantum Decoherence Effects in Neutrino Oscillations at DUNE}


\author{G. Balieiro Gomes}
\email[]{balieiro@ifi.unicamp.br}
\affiliation{Instituto  de  F\'isica  Gleb  Wataghin\\  Universidade  Estadual  de  Campinas - UNICAMP\\ Rua S\'ergio Buarque de Holanda, 777 \\  13083-970,  Campinas,  S\~ao Paulo,  Brazil}

\author{D. V. Forero}
\email[]{dvanegas@ifi.unicamp.br}
\affiliation{Instituto  de  F\'isica  Gleb  Wataghin\\  Universidade  Estadual  de  Campinas - UNICAMP\\ Rua S\'ergio Buarque de Holanda, 777 \\  13083-970,  Campinas,  S\~ao Paulo,  Brazil}

\author{M. M. Guzzo}
\email{guzzo@ifi.unicamp.br}
\affiliation{Instituto  de  F\'isica  Gleb  Wataghin\\  Universidade  Estadual  de  Campinas - UNICAMP\\ Rua S\'ergio Buarque de Holanda, 777 \\  13083-970,  Campinas,  S\~ao Paulo,  Brazil}

\author{P. C. de Holanda}
\email{holanda@ifi.unicamp.br}
\affiliation{Instituto  de  F\'isica  Gleb  Wataghin\\  Universidade  Estadual  de  Campinas - UNICAMP\\ Rua S\'ergio Buarque de Holanda, 777 \\  13083-970,  Campinas,  S\~ao Paulo,  Brazil}

\author{R. L. N. Oliveira}
\email{robertol@ifi.unicamp.br}
\affiliation{Instituto  de  F\'isica  Gleb  Wataghin\\  Universidade  Estadual  de  Campinas - UNICAMP\\ Rua S\'ergio Buarque de Holanda, 777 \\  13083-970,  Campinas,  S\~ao Paulo,  Brazil}
\affiliation{Universidade Federal do ABC - UFABC, Santo Andr\'e, S\~ao Paulo, Brazil}


\date{\today}

\begin{abstract}
In this work we analyze quantum decoherence in neutrino oscillations considering the Open Quantum System framework and oscillations through matter for three neutrino families. Taking DUNE as a case study we performed sensitivity analyses for two neutrino flux configurations finding limits for the decoherence parameters. We also offer a physical interpretation for a new peak which arises at the $\nu_{e}$ appearance probability with decoherence. The sensitivity limits found for the decoherence parameters are $\Gamma_{21}\le 1.2\times10^{-23}\,\text{GeV}$ and $\Gamma_{32}\le 7.7\times10^{-25}\,\text{GeV}$ at $90\%$ C. L. 
\end{abstract}


\maketitle


\section{Introduction}

Even though the standard three neutrino oscillation paradigm is well established and several oscillation parameters have been 
already measured with certain precision~\cite{revdavid}, the quest for establishing the violation of the Charge Parity (CP) 
symmetry in the leptonic sector, the octant preference or the maximality of the atmospheric mixing angle, and the neutrino mass 
ordering is still ongoing. In order to fulfill such goals and also to reach a greater precision in the measurement of all the 
neutrino oscillation parameters, future experiments such as the Deep Underground Neutrino Experiment (DUNE) 
\cite{cdr1,cdr2,cdr3,cdr4,annex3a} are being developed. DUNE is a long-baseline neutrino experiment where the neutrinos produced 
at Fermilab are detected at the Sanford Underground Research Laboratory, therefore after traveling $\sim 1300\,\text{km}$. DUNE is 
designed to study the $\nu_\mu$ and $\nu_{e}$ (and also $\bar{\nu}_\mu$ and $\bar{\nu}_e$) oscillations through the Earth crust's 
matter, and it is expected to provide a measurement of the neutrino mass hierarchy. DUNE is also sensitive to the Dirac phase 
present in the lepton mixing matrix, which parameterizes the possibility that neutrinos violate the CP symmetry. In order to 
perform these major discoveries and the precise measurement of the atmospheric mixing angle, DUNE will have to reach a novel 
control of systematics and very large statistics. Such features can be used not only to achieve the main goals for the standard 
oscillation program, but more importantly, can also be useful to probe new physics effects, such as decoherence.

There are several works 
\cite{robertot13,robertosun,balikamland,lisi,roberto,coelhoprl,robertodune,coelhopeak,robertominos,Carpio:2017nui, 
	nunokawadec} showing how decoherence can emerge in models considering interactions between a neutrino subsystem and an 
environment in the Open Quantum System \cite{breuer} framework, and some of these works present analyses of possible 
constraints for the decoherence parameters \cite{balikamland,robertominos, nunokawadec}. Nevertheless, there are other 
experiments which could be considered and might be suitable to make a full three neutrino family analysis. As will be 
shown later on, the decoherence effect arises in the oscillation probabilities through damping terms depending on the 
baseline, suggesting that a long baseline experiment such as DUNE is an excellent candidate to bound all the decoherence 
parameters for three neutrino families. Although it is speculated that the quantum decoherence effect could be generated 
by quantum gravity \cite{ellis}, in this work we will use a phenomenological approach. We do not use any microscopical 
model which describes the source of such effects, and therefore such hypothesis or other possible origins of decoherence 
will not be discussed. It is also important to point out that in this work we will study only the decoherence effects 
which arise in the framework of Open Quantum Systems, we will not address decoherence effects from wave packet 
separation (see for example Refs. \cite{wavedec, dayawavedec}), which are already present within usual Quantum 
Mechanics.

This work is organized in the following way. We review how one can study neutrino oscillations considering a coupling with the 
environment in the Quantum Open System framework in Section~\ref{formalism}, presenting also the form of the oscillation 
probabilities with decoherence in three families. In Section~\ref{sec:probability} we offer a physical interpretation of a new 
peak that arises in the oscillation probabilities with decoherence. In Section~\ref{results} we perform sensitivity analyses and 
present the sensitivity regions found for the decoherence parameters. Since the optimized flux configuration at DUNE already 
covers a broad range of neutrino energies, DUNE is sensitive to the decoherence parameters. We also consider a high energy flux 
configuration to reach the high energy peak induced by decoherence in the appearance channel, which is the `smoking gun' for 
decoherence, providing increased sensitivity to the decoherence parameter $\Gamma_{32}$.

\section{Formalism} \label{formalism}
When the coupling between the neutrino subsystem and the environment is considered, the time evolution of the subsystem density operator $\rho$ is given by the Lindblad Master Equation \cite{roberto}:

\begin{equation}
\frac{d}{dt} \rho (t) = L \rho = -i [H,\rho] + \frac{1}{2} \displaystyle \sum_{k=1}^{N^2-1} \left( [ V_k , \rho V_k^\dag ] + [ V_k \rho , V_k^\dag ] \right),
\label{lindblad}
\end{equation} where, $H$ is the subsystem's Hamiltonian, $V_k$ are the operators responsible for the interactions between the subsystem and the environment, and \textit{N} is the dimension of the Hilbert space of the subsystem.
The non-Hamiltonian term can be written as:

\begin{equation}
D [\rho (t) ] = \frac{1}{2} \displaystyle \sum_{k=1}^{N^2-1} \left( [ V_k , \rho V_k^\dag ] + [ V_k \rho , V_k^\dag ] \right),
\end{equation} which will be referred from now on as dissipator.

The matrix $D$ is subjected to constraints to assure that the operator $\rho (t)$ have all the properties of a density operator and that its physical interpretation is correct. In particular it can be shown that the operator $V$ must be hermitian ($V_k = V_k^\dag$)~\cite{robertot13}, to ensure that the system's entropy increases in time. 

In the case of three active neutrinos one can expand the elements on Lindblad equation in Eq.~(\ref{lindblad}) using the $SU(3)$ generators, the Gell-Mann matrices $\lambda_i$, as a basis:
\[
H=H_i\lambda_i~~~;~~~\rho=\rho_j\lambda_j
\]
where the sum over repeated indices are implied, and Eq..~(\ref{lindblad}) can be rewritten as:
\begin{equation}
\frac{d}{dt} \rho_k (t) \lambda_k = f^{ijk} H_i \rho_j (t) \lambda_k  + D_{kl} \rho_l \lambda_k,
\label{lindbladsu3}
\end{equation} where the $f^{ijk}$ are structure constants completely antisymmetric in the indices $i,j,k$.

We assume $D_{kl}$ as a symmetric matrix and with $ D_{k 0 } = D_{0 l} = 0$ in order to have probability conservation. We will 
also impose that $[H,V_{k}]=0$, which implies energy conservation in the neutrino subsystem. Other conditions for $D$ will come 
from the imposition that it satisfies the criteria for complete positivity, which must be obeyed by a density operator, and hence 
also by the dissipator. For three neutrino families these criteria are described in Ref.~\cite{robertot13} and references 
therein. Although the derivation of these conditions can be found in these references, we found it worthwhile to 
present them again in Appendix~\ref{appendixA} for the specific case analyzed here, namely, with energy conservation in the 
neutrino sector.

Under such constraints the dissipative matrix $D_{k l}$ assumes the following form:
\begin{equation}
D_{kl}=-{\rm diag}\{\Gamma_{21},\Gamma_{21},0,\Gamma_{31},\Gamma_{31},\Gamma_{32},\Gamma_{32},0\}.
\label{dissipdune}
\end{equation}

The decoherence parameters are not independent from each other, and are related by the following equations~\cite{robertodune}:

\begin{equation}
\Gamma_{21} = 2 a_{3}^{2}\geq0;
\label{g21eq}
\end{equation}

\begin{equation}
\Gamma_{31} = \frac{1}{2} (a_{3}+a_{8})^{2}\geq0;
\label{g31eq}
\end{equation}

\begin{equation}
\Gamma_{32} = \frac{1}{2} (a_{3} - a_{8})^{2}\geq0;
\label{g32eq}
\end{equation} where the $a_i$ are the terms of the expansion of the $V_k$ operators in terms of the SU(3) matrix basis:
\begin{equation}
V_k=a_n^k \lambda_n.
\label{vkeq}
\end{equation}

Since we have that a density matrix must be positive semi-definite, which means that if $\lambda_i$ are its eigenvalues, 
then $\lambda_i \geq 0 \; \forall i$ \cite{cohen}, it is clear that the dissipator in Eq.~(\ref{dissipdune}) with the 
conditions in Eqs.~(\ref{g21eq}) -- (\ref{g32eq}) satisfies the needed criteria in order to preserve its physical 
meaning. In the appendix~\ref{appendix} we discuss the validity of the dissipator in Eq.~(\ref{dissipdune}) and the 
positivity conditions when one considers decoherence in vacuum or in constant density matter, highlighting the 
differences between our approach and the one used in Ref.~\cite{Carpio:2017nui}. In the following sections we consider 
$\Gamma_{21}$ and $\Gamma_{32}$ as the independent parameters, and $\Gamma_{31}$ given by equations~(\ref{g21eq}) - 
(\ref{g32eq}).

Considering DUNE baseline, matter effects have to be taken into account. The complete Hamiltonian in the flavor basis is then given by:

\begin{equation}
H = \left[ U \left( \begin{array}{ccc}
0 & 0 & 0  \\
0 & \frac{\Delta m_{21}^2}{2E} & 0  \\
0 & 0 & \frac{\Delta m_{31}^2}{2E} \\

\end{array} \right) U^{\dagger} +  \left( \begin{array}{ccc}
\hat{A} & 0 & 0  \\
0 & 0 & 0  \\
0 & 0 & 0 \\

\end{array} \right)
\right],
\label{ham3fdune}
\end{equation} where the $\Delta m_{ij}^2 \equiv m_i^2 - m_j^2$  are the squared mass differences between the mass eigenstates and $E$ is the neutrino energy $\hat{A} = \sqrt{2} G_F n_e$ is the matter potential where $G_F$ is the Fermi coupling constant and  $n_e$ is the electron number density, and $U$ is the mixing matrix for three neutrino families, which is given by:

\begin{widetext}
\begin{eqnarray}
U & = & \left( \begin{array}{ccc}
c_{12}c_{13} & s_{12}c_{13} & s_{13} e^{-i \delta_{CP}}  \\
-s_{12}c_{23} - c_{12}s_{23}s_{13}e^{i \delta_{CP}} & c_{12}c_{23} -  s_{12}s_{23}s_{13}e^{i \delta_{CP}} & s_{23}c_{13}  \\
s_{12}s_{23} - c_{12}c_{23}s_{13}e^{i \delta_{CP}} & - c_{12}s_{23} - s_{12}c_{23}s_{13}e^{i \delta_{CP}} & c_{23}c_{13} \\
\end{array} \right),
\label{pmns}
\end{eqnarray} \end{widetext} and where $c_{ij}$ and $s_{ij}$ denote $\cos(\theta_{ij})$ and $\sin(\theta_{ij})$ respectively.

The Eq.~(\ref{lindbladsu3}) will be solved in the effective mass eigenstate basis, hence we must find the diagonal form of the Hamiltonian:

\begin{equation}
H =  \frac{1}{2 E} \left( \begin{array}{ccc}
0 & 0 & 0  \\
0 & \tilde{\Delta}_{21} & 0  \\
0 & 0 & \tilde{\Delta}_{31}\\
\end{array} \right),
\label{ham3fdiag}
\end{equation} where $\Delta_{ij}$ are the effective squared mass differences of neutrinos in matter.


Solving the Lindblad equation, with the dissipator defined in Eq.~(\ref{dissipdune}), one finds:
\begin{widetext}
\begin{equation}
\rho_{\tilde{m}}(x) = 
\left(\begin{array}{c c c } 
\rho_{11}(0) & \rho_{12}(0)e^{-(\Gamma_{21} +i\tilde{\Delta}_{21})^{*}x} & \rho_{13}(0)e^{-(\Gamma_{31} +i\tilde{\Delta}_{31})^{*}x} \\
\rho_{21}(0)e^{-(\Gamma_{21} +i\tilde{\Delta}_{21})x} &\rho_{22}(0) & \rho_{23}(0)e^{-(\Gamma_{32} +i\tilde{\Delta}_{32})^{*}x} \\
\rho_{31}(0)e^{-(\Gamma_{31} +i\tilde{\Delta}_{31})x} & \rho_{32}(0)e^{-(\Gamma_{32} +i\tilde{\Delta}_{32})x} &\rho_{33}(0)  \end{array} \right),
\label{rhoxdune}
\end{equation} \end{widetext} where $\rho_{ij}(0)$ are the elements of the density matrix for the initial state.

The oscillation probabilities for each channel can be calculated from:

\begin{equation}
P_{\nu_\alpha \nu_{\alpha^{\prime}} }  = Tr[\rho_{\alpha}(0)\rho_{\alpha^{\prime}}(x)] \,.
\label{tracedune}
\end{equation}


Using Eq.~(\ref{rhoxdune}) and Eq.~(\ref{tracedune}), and after some algebraic manipulation, one obtains:

\begin{equation}
\begin{split}
P_{\nu_\alpha \nu_{\alpha^{\prime}} }= & \delta_{\alpha \alpha^{\prime}} - 2 \sum_{j>k} Re(\tilde{U}_{\alpha^{\prime} j} \tilde{U}_{\alpha j}^{\ast} \tilde{U}_{\alpha k} \tilde{U}_{\alpha^{\prime} k}^{\ast}) \\
& + 2 \sum_{j>k} Re(\tilde{U}_{\alpha^{\prime} j} \tilde{U}_{\alpha j}^{\ast} \tilde{U}_{\alpha k} \tilde{U}_{\alpha^{\prime} k}^{\ast}) e^{- \Gamma_{jk} x} \cos \left( \frac{\tilde{\Delta}_{jk}}{2E} x \right) \\
& + 2  \sum_{j>k} Im(\tilde{U}_{\alpha^{\prime} j} \tilde{U}_{\alpha j}^{\ast} \tilde{U}_{\alpha k} \tilde{U}_{\alpha^{\prime} k}^{\ast}) e^{- \Gamma_{jk} x} \sin\left( \frac{\tilde{\Delta}_{jk}}{2E} x \right),
\end{split}
\label{probdune}
\end{equation} where $\tilde{U}$ is the unitary mixing matrix which diagonalizes the Hamiltonian in the presence of matter effects. To obtain the corresponding probability for antineutrinos one must repeat the procedure above changing $\hat{A} \rightarrow - \hat{A}$ in Eq.~(\ref{ham3fdune}) and $\delta_{CP} \rightarrow -\delta_{CP}$ in Eq.~(\ref{pmns}). It is also important to point out that we assumed the decoherence parameters $\Gamma_{jk}$ as being equal for both neutrinos and antineutrinos, differently from what is done by Ref.~\cite{gbarenboim} where  CPT violation in quantum decoherence is used to fit the LSND oscillation data without the inclusion of sterile neutrinos.

In the following sections we present results from the implementation of Eq.~(\ref{probdune}) in a modified version of the GLoBES~\cite{globes1,globes2} probability engine, which was also double-checked by solving numerically the Lindblad Equation in Eq.~(\ref{lindbladsu3}).

\section{Effects of Decoherence on the Oscillation Probabilities}\label{sec:probability}

We consider the four oscillation channels, appearance and disappearance for both neutrino and antineutrino modes, for benchmark values of the decoherence parameters $\Gamma_{21}, \Gamma_{31}, \Gamma_{32}$. For the probability studies, only the DUNE baseline ($L=1300~\text{km}$) and its energy range (which 
extends from hundreds of MeV's to tenths of GeV's) are needed. The values of the standard oscillation parameters used along this work are given in Table~\ref{eq:parameters-val}.


\begin{table}[!htb]
	\centering
	 	\begin{tabular*}{\columnwidth}{@{\extracolsep{\fill}} c c}
	 		\hline \hline
	 		$\sin^2{\theta_{12}}$ & $0.321$\\ \hline
	 		$\sin^2{(2 \theta_{13})}$ & $0.0841$ \\ \hline
	 		$\sin^2{(2\theta_{23})}$ & $0.99$ \\ \hline
	 		$\delta_{CP}$ & $-\pi/2$ \\ \hline
	 		$\Delta m^2_{21}$ & $7.56\times10^{-5} \,\text{eV}^2$ \\ \hline
	 		$\Delta m^2_{31}$ & $2.55\times10^{-3} \,\text{eV}^2$ \\ \hline\hline
	 	\end{tabular*}
	\caption{Values for the standard oscillation parameters from Refs. \cite{revdavid,dayabay017,t2k0172}.}
	\label{eq:parameters-val}
\end{table}

\begin{figure*}[!htb]
	\centering
	\subfloat[$\nu_e $ appearance]{{\includegraphics [scale=0.3]{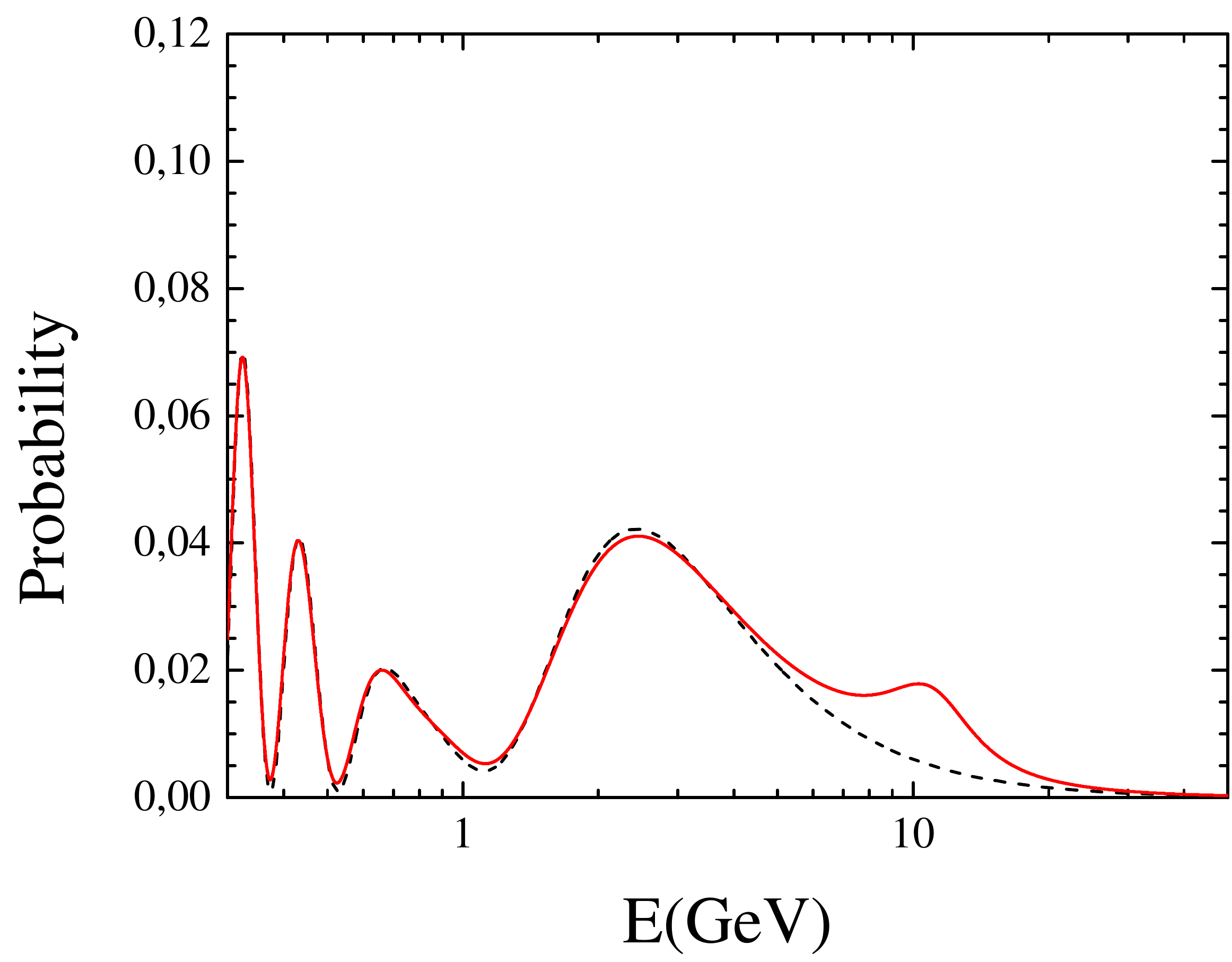}}}%
	\qquad
	\subfloat[$\bar{\nu}_e $ appearance]{{\includegraphics [scale=0.3]{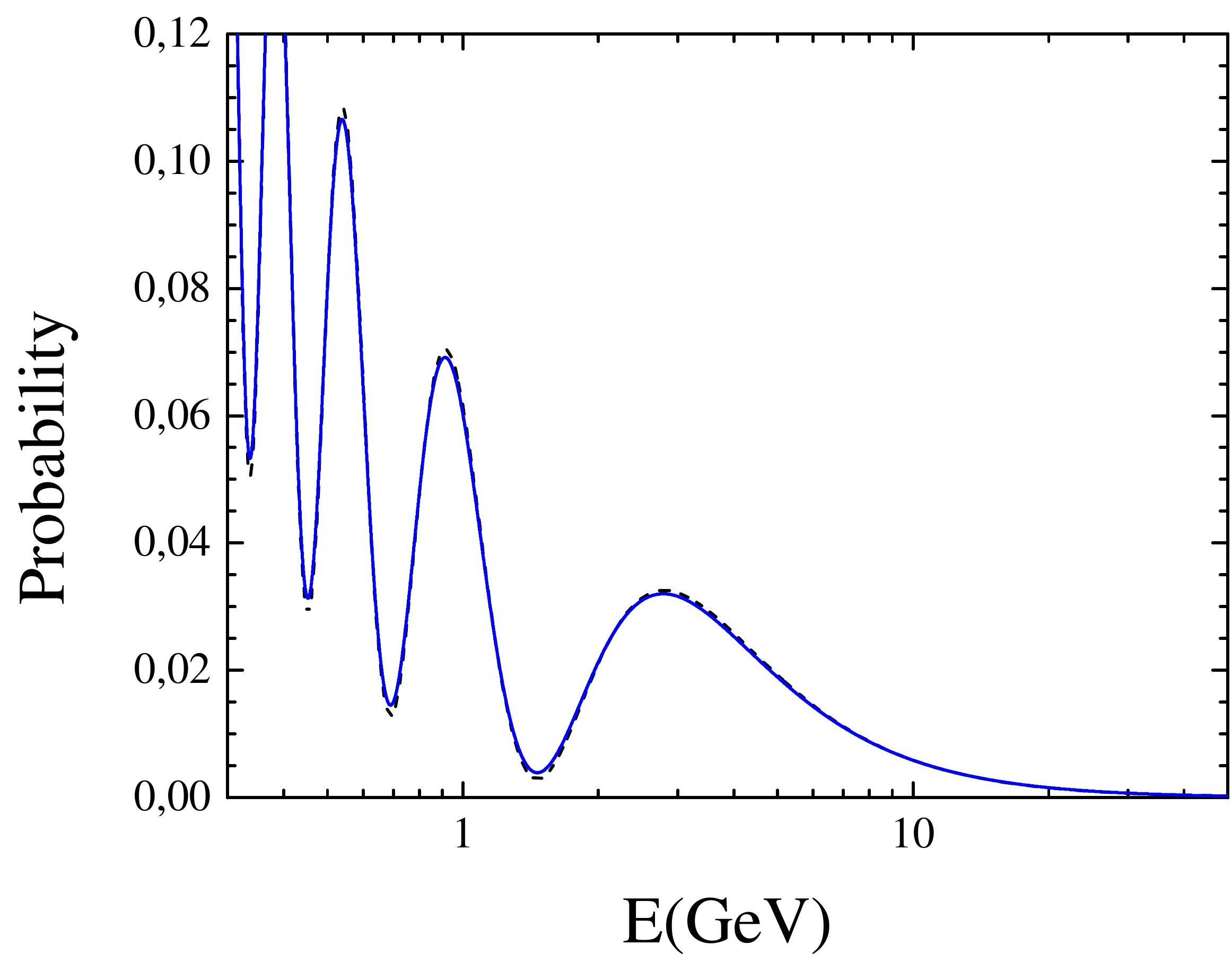}}}%
	\qquad
	\subfloat[$\nu_\mu $ disappearance]{{\includegraphics [scale=0.3]{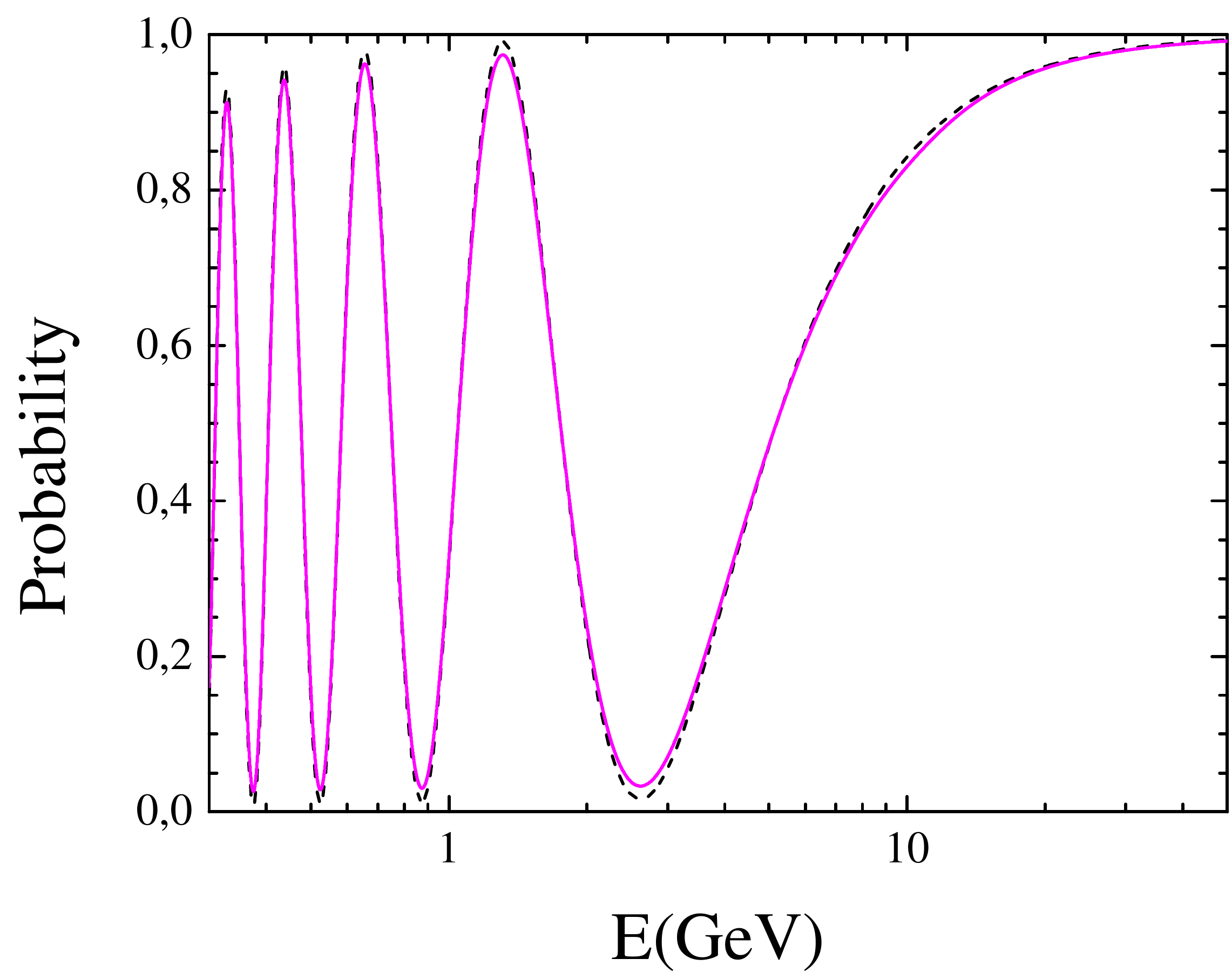}}}%
	\qquad
	\subfloat[$\bar{\nu}_\mu $ disappearance]{{\includegraphics [scale=0.3]{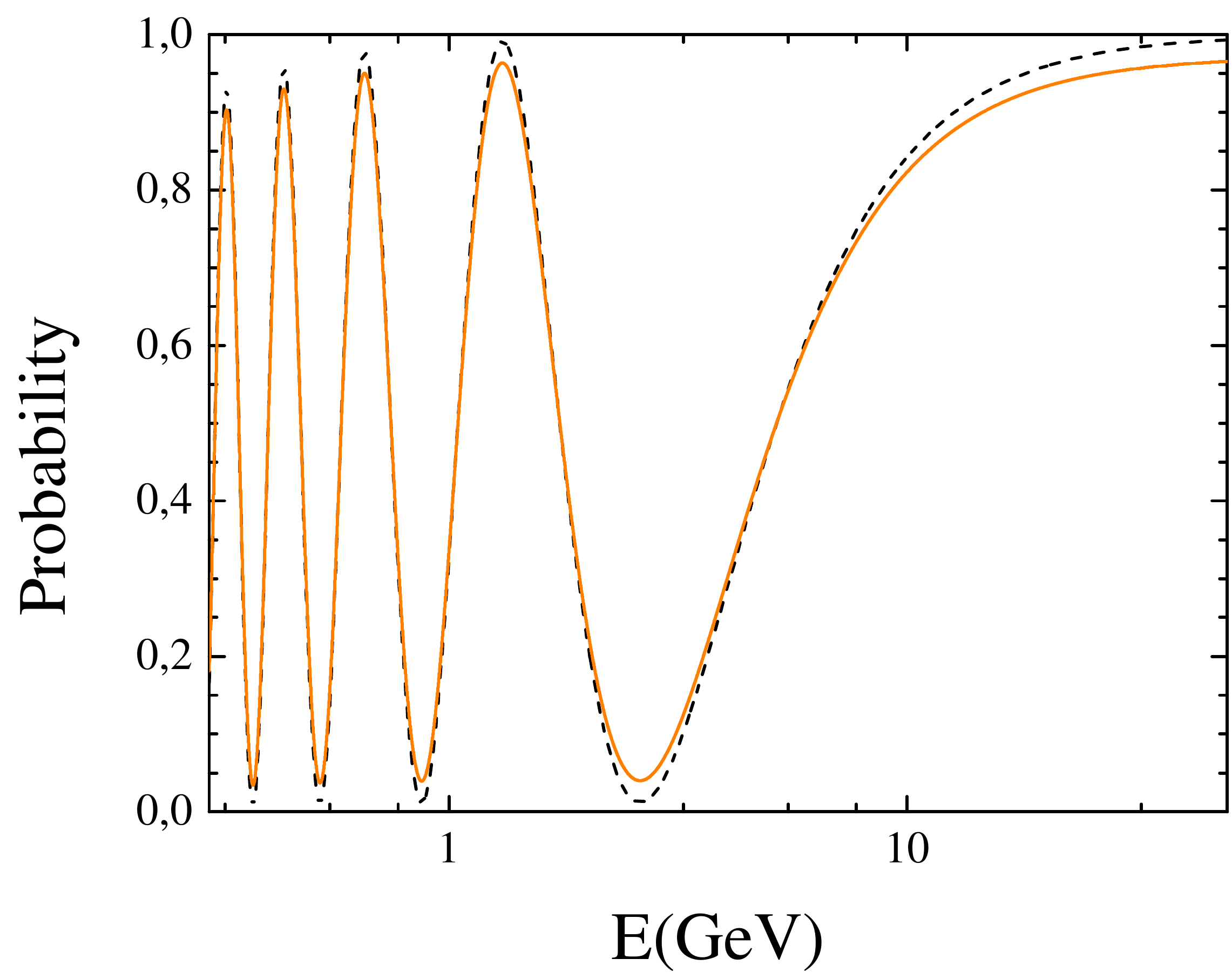}}}%
	
	\caption{Oscillation probabilities using: $\Gamma_{21} = 5.1 \times 10^{-25}$ GeV,  $\Gamma_{32} = 8.9 \times 10^{-24}$ GeV (solid line) and Standard (dashed line). The values of the oscillation parameters were set according to Table~\ref{eq:parameters-val}, and $\Gamma_{31}$ was calculated according to Eqs.~(\ref{g21eq}) - (\ref{g32eq}).}
	\label{4prob1}
\end{figure*}

As we can see in Fig.~\ref{4prob1}, the decoherence parameters affect the four oscillation channels, and for the values of the decoherence parameters considered we can see a few different effects on the oscillation probabilities. In Figs.~\ref{4prob1} (c) and (d) there is a small decrease in the overall oscillation amplitude (more accentuated for the $\bar{\nu}_\mu $ disappearance probability). In Fig.~\ref{4prob1} (d) we can also see a decrease in the $\bar{\nu}_\mu $ for $E\gtrsim 10$ GeV. However, the most striking difference respect to the standard oscillation is the new peak at $\sim 10$ GeV for the $\nu_{e}$ appearance probability in the presence of decoherence, which would provide a clear signature of new physics. In the next section we will discuss this feature in more details. Although the peak by itself is not a novelty, and it was somehow studied in previous works (see for instance~\cite{robertodune,coelhopeak}), here we provide a detailed physical interpretation, and more importantly, we suggest how this unique feature of decoherence can be probed at DUNE.

\subsection{New peak at the $\nu_{e}$ appearance probability: physical interpretation}\label{sec:peak}

\begin{figure}[!htb]
	\centering
	\includegraphics [width=0.45\textwidth]{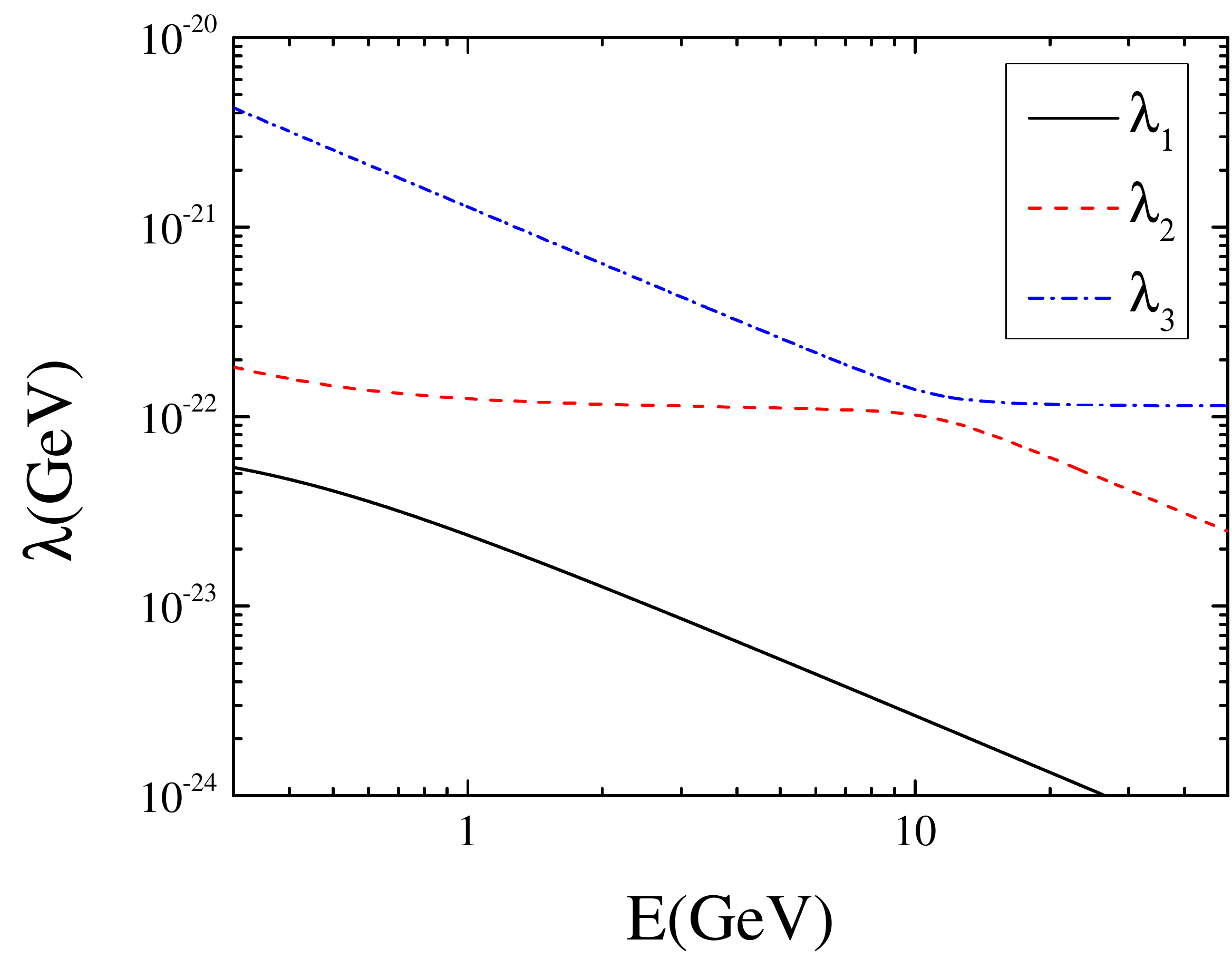}
	\caption{\label{evfig} Eigenvalues ($\lambda_i$, i=1,2,3) of the Hamiltonian in Eq.~(\ref{ham3fdune}). We can see an indication of a resonance region about $E \sim 10$ GeV.}
\end{figure}

A peak at $\sim 10$ GeV is present in the $\nu_{e}$ appearance probability in the presence of decoherence. In order to obtain a physical insight of this new feature, let us begin by analysing the behavior of the eigenvalues of the Hamiltonian ($\lambda$) in Eq.~(\ref{ham3fdune}), which can be seen in Fig.~\ref{evfig}.

As we can see in Fig.~\ref{evfig}, there is a level crossing between the eigenvalues referred as 2 and 3 at $E \sim 10$ GeV, which indicates a resonance at that energy for the parameters considered.

From the oscillation probabilities with decoherence in Eq.~(\ref{probdune}), the $\Gamma_{jk}$ parameters appear in the form of $e^{- \Gamma_{jk} x}$ damping factors for the terms:

\begin{equation}
\begin{split}
I_{\alpha \alpha^{\prime} } = & 2 \sum_{j>k} Re(\tilde{U}_{\alpha^{\prime} j} \tilde{U}_{\alpha j}^{\ast} \tilde{U}_{\alpha k} \tilde{U}_{\alpha^{\prime} k}^{\ast}) \cos \left( \frac{\tilde{\Delta}_{jk}}{2E} x \right) \\
& + 2  \sum_{j>k} Im(\tilde{U}_{\alpha^{\prime} j} \tilde{U}_{\alpha j}^{\ast} \tilde{U}_{\alpha k} \tilde{U}_{\alpha^{\prime} k}^{\ast})  \sin \left( \frac{\tilde{\Delta}_{jk}}{2E} x \right),
\end{split}
\label{interferenceeq}
\end{equation} for $j,k = 1,2,3$ and $j>k$.

\begin{figure}[!htb]
	\centering
	\includegraphics [width=0.4\textwidth]{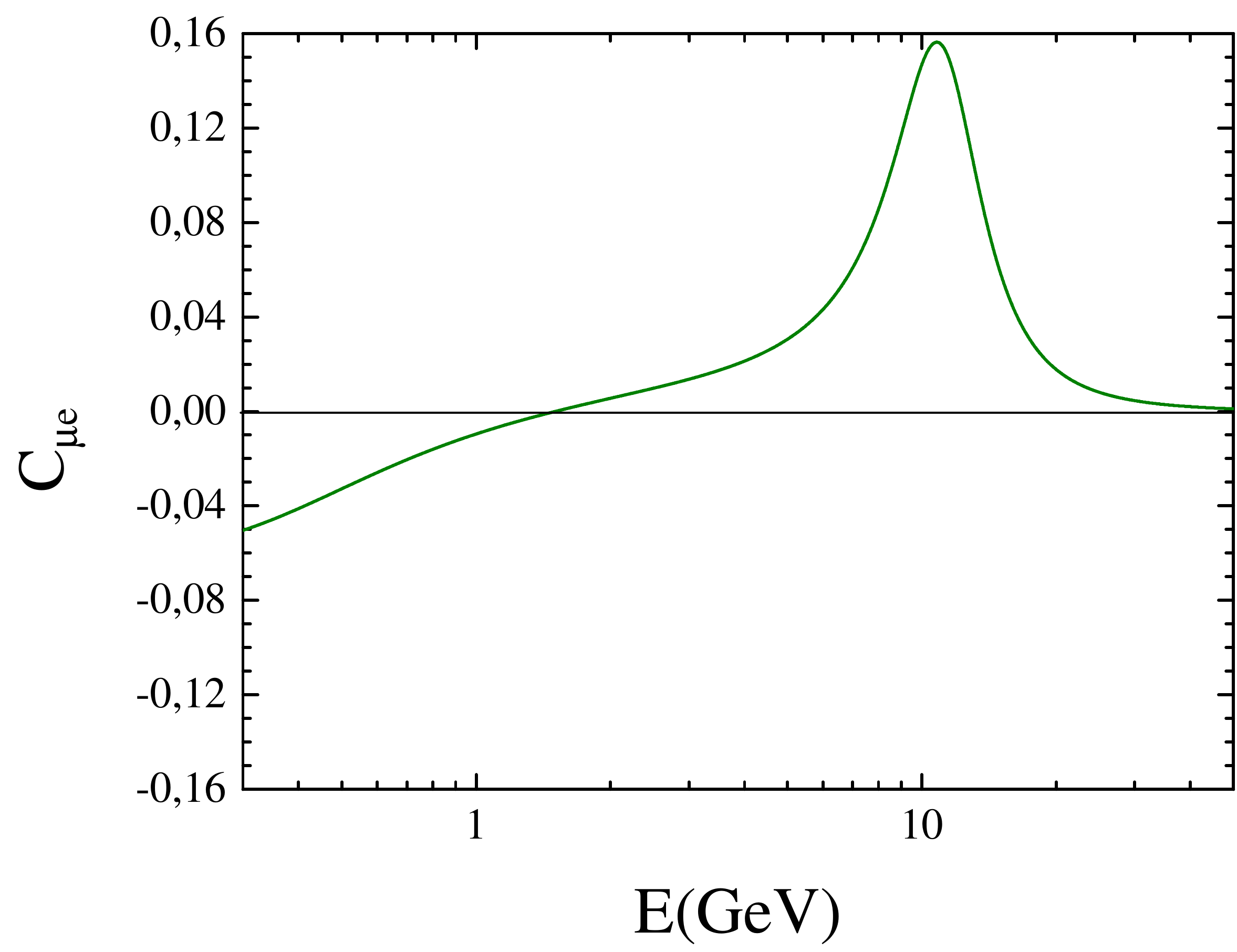}
	\includegraphics [width=0.4\textwidth]{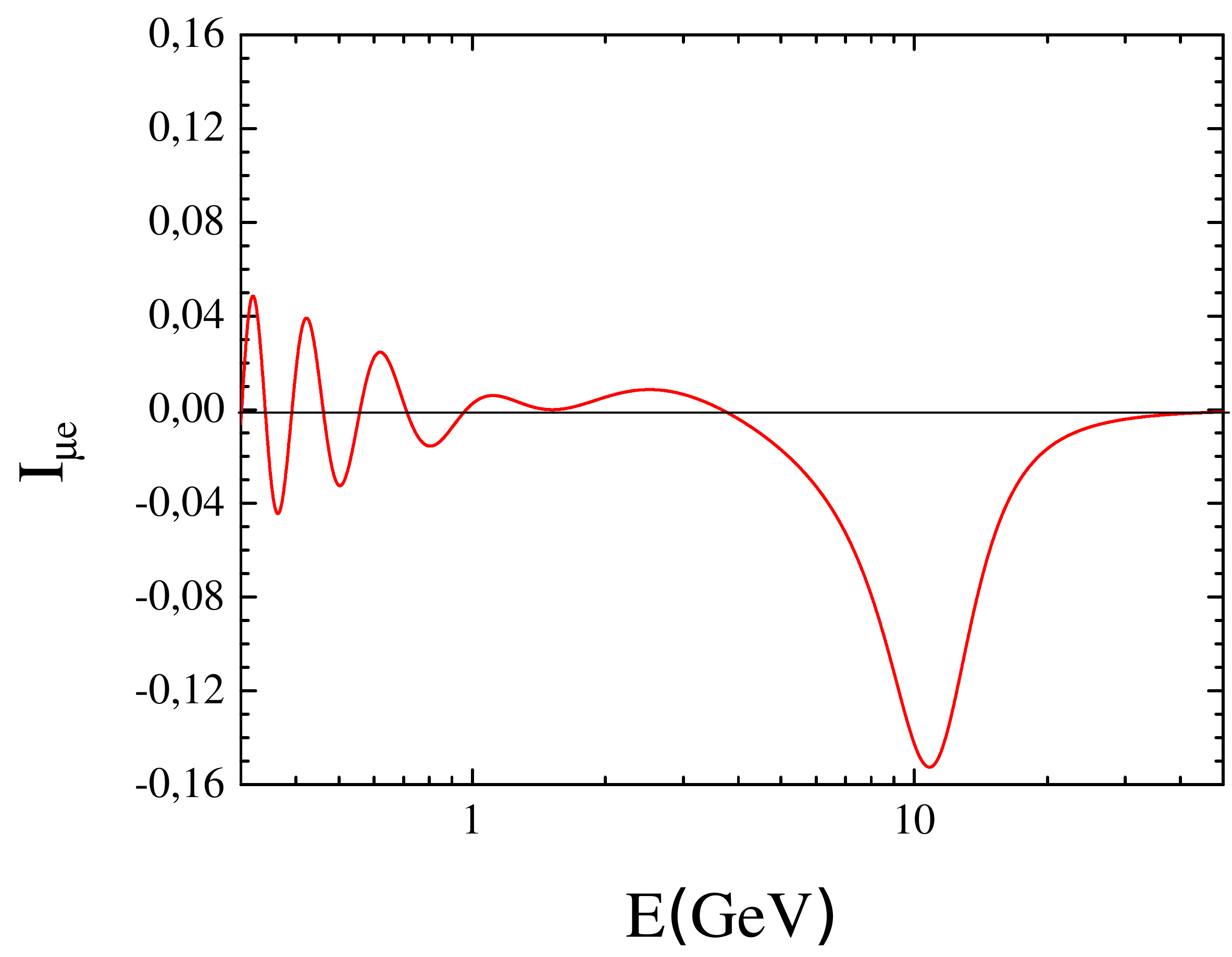}
	\caption{\label{interferencefig} Left panel: behavior of $C_{\mu e }$ for $j=3$, $k=2$, $\alpha=\mu$, $\alpha^{\prime} = e$ given by Eq.~(\ref{32const}). Right panel: interference factor $I_{\mu e }$ for $j=3$, $k=2$, $\alpha=\mu$, $\alpha^{\prime} = e$ given by Eq.~(\ref{32interf}).}
\end{figure}

Since $I_{\alpha \alpha^{\prime} }$ is the term of the probability where we have the dependence on the oscillation phase through $\cos\left( \frac{\tilde{\Delta}_{jk}}{2E} x \right)$ and $\sin\left( \frac{\tilde{\Delta}_{jk}}{2E} x \right)$, which are responsible for the quantum interference in the oscillation probabilities, we will refer to it as the {\it interference factor}. In addition, there are terms not affected by the decoherence parameters:

\begin{equation}
C_{\alpha \alpha^{\prime} } = \delta_{\alpha \alpha^{\prime}} - 2 \sum_{j>k} Re(\tilde{U}_{\alpha^{\prime} j} \tilde{U}_{\alpha j}^{\ast} \tilde{U}_{\alpha k} \tilde{U}_{\alpha^{\prime} k}^{\ast}),
\label{contanteq} 
\end{equation} where $j,k = 1,2,3$ and $j>k$. The term $C_{\alpha \alpha^{\prime} }$ in the case of the $\nu_e$ appearance probability, and for $j=3$ and $k=2$, is given by:

\begin{equation}
C_{\mu e } =  -2~ Re(\tilde{U}_{e 3} \tilde{U}_{\mu 3}^{\ast} \tilde{U}_{\mu 2} \tilde{U}_{e 2}^{\ast}),
\label{32const}
\end{equation} 
is also shown in the left panel of Fig.~\ref{interferencefig}. This term of the $\nu_e$ appearance probability indeed presents a resonance at $E \sim 10$ GeV, which was already suggested by the level crossing in the eigenvalues for this energy in Fig.~\ref{evfig}.

Let us now analyse the effect of $e^{-\Gamma_{32}x}$ over the $\nu_e$ appearance probability, and in order to do so, we considered the form of the interference factor in Eq.~(\ref{interferenceeq}) when $j=3$, $k=2$, $\alpha=\mu$, $\alpha^{\prime} = e$:

\begin{equation}
\begin{split}
I_{\mu e } = & ~2~ Re(\tilde{U}_{e 3} \tilde{U}_{\mu 3}^{\ast} \tilde{U}_{\mu 2} \tilde{U}_{e 2}^{\ast}) \cos \left( \frac{\tilde{\Delta}_{32}}{2E} x \right) \\
& + 2~    Im(\tilde{U}_{e 3} \tilde{U}_{\mu 3}^{\ast} \tilde{U}_{\mu 2} \tilde{U}_{e 2}^{\ast}) \sin \left( \frac{\tilde{\Delta}_{32}}{2E} x \right).
\end{split}
\label{32interf}
\end{equation}

The form of the interference term, which is subject to the damping factor $e^{-\Gamma_{32}x}$, and as given in Eq.~(\ref{32interf}), is shown in the right panel of Fig.~\ref{interferencefig}. We can see that the term $I_{\mu e }$ work as an interference factor to the oscillation probabilities, in particular, we see that around $E = 10.8$ GeV, which is exactly the energy of the resonance, we have a strong destructive interference. For the standard oscillation probabilities (without decoherence), such destructive interference would exactly cancel out the resonance at $E \sim 10.8$ Gev shown in the left panel. In fact, considering a constant matter density of $2.96 g/cm^3$ \cite{prem1,prem2}, numerically the maximum of $C_{\mu e }$ in Eq.~(\ref{32const}) is equal to the minimum of $I_{\mu e }$ in Eq.~(\ref{32interf}), and both coincide at $E = 10.82$ GeV. However, when we have oscillations with decoherence the term $e^{- \Gamma_{32} x}$ work as a damping to this interference factor, therefore eliminating the destructive interference at $E \sim 10$ GeV. The elimination of such destructive interference enhances the $\nu_e$ appearance probability, since now the destructive interference cannot completely cancel out the resonance, therefore creating the peak shown in Fig.~\ref{4prob1} (a). Since such peak constitutes a very significant effect in the oscillation probabilities in the presence of decoherence, being able to reconstruct it will provide a compelling test of decoherence. If DUNE is compatible with standard oscillations, severe bounds to the decoherence parameters can be obtained as long as the experiment measure a significant number of events around $E \sim 10$ GeV.

\section{Results}\label{results}

In this section we are going to show sensitivity analyses considering neutrino oscillations with decoherence in matter given by Eq.~(\ref{probdune}). We first show how each oscillation channel is sensible to decoherence by calculating the event rates, and then we establish DUNE sensitivities to the decoherence parameters. For the sensitivity analysis we have considered two neutrino flux configurations, as will be detailed in the following sections, to exploit the main features of the decoherence effects discussed previously.

In the following studies, we assume the DUNE configuration as defined in the CDR document in Ref.~\cite{cdr2} and in particular we 
made use of the GLoBES files from Ref.~\cite{ancillary}. Basically, it is assumed DUNE will be running for $3.5$ years in each 
mode (neutrino and antineutrino), a fiducial mass of the far detector (liquid Argon) of $40\,\text{kt}$, and the default flux beam 
power of $1.07\, \text{MW}$. The channels considered in each analysis are defined for each case. The systematical errors, energy 
resolution, and efficiencies are fixed to the values in the CDR studies.

\subsection{Relative events with the DUNE default flux configuration} \label{sec:relr}

For a particular input for the decoherence parameters, the total number of events and the energy event spectra are calculated for each oscillation channel. We define the relative Event Rates as $\delta R_{rel}$:

\begin{equation}
\delta R_{rel} = \frac{R(\Gamma_{ij}\neq 0) - R(\Gamma_{21}=\Gamma_{31}=\Gamma_{32}=0)}{R(\Gamma_{21}=\Gamma_{31}=\Gamma_{32}=0)},
\label{deltar}
\end{equation} 
where$\Gamma_{ij}\neq 0$ (\textit{i.e.}: $\Gamma_{21}\neq 0, \Gamma_{31} \neq 0, \Gamma_{32} \neq 0$) are chosen in order to satisfy Eqs.~(\ref{g21eq}) -- (\ref{g32eq}), and $R$ correspond to the event rates.

\begin{figure}[!htb]
	\centering
	\includegraphics[height=0.35\textwidth]{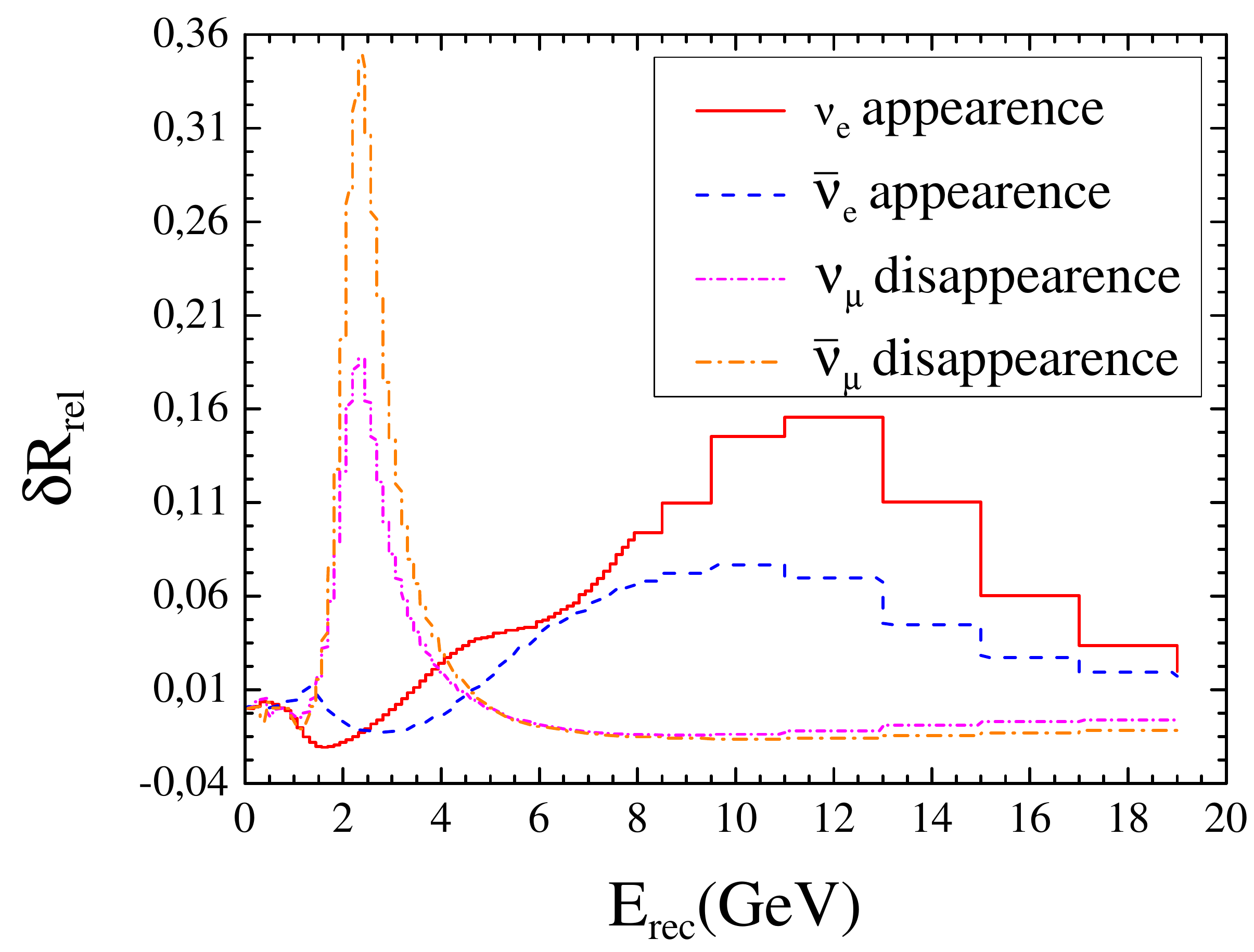}
	\caption{Relative event rates defined in Eq.~(\ref{deltar}) setting $\Gamma_{21} = 5.1 \times 10^{-25}$ GeV, and $\Gamma_{32} = 8.9 \times 10^{-24}$ GeV, for the event rates in the presence of decoherence.}
	\label{4prob3}
\end{figure}

Fig.~\ref{4prob3} show the relative deviation of the number of 
$\nu_e$, $\bar{\nu}_e$, $\nu_\mu$, and $\bar{\nu}_\mu$ events in respect to the standard oscillation case 
without decoherence. From the $\nu_e$ and 
$\bar{\nu}_e$ events, one can see a low relative deviation ($ < 3\%$) at the DUNE flux 
(default) maximum ($\sim 2.5\,\text{GeV}$). The peak at $E \gtrsim 10$ GeV in the $\nu_e$ ($\bar{\nu}_e$) events is also 
relatively low, being about $\sim 16\%$ ($\sim 8\%$), but this is expected because with the default flux events at the high 
energy end of the spectrum are much smaller than in the DUNE energy peak. In the case of $\nu_\mu$ and $\bar{\nu}_\mu$ events in 
Fig.~\ref{4prob3} we can notice that at slightly lower energies from the DUNE 
flux (default) maximum, a relative deviation of the order of $\sim 19\%$ is obtained in 
the case of $\nu_\mu$ and $\sim 35\%$ for $\bar{\nu}_\mu$ events.

It appears to be that, with the default flux configuration, DUNE is sensitive to decoherence, and this 
sensitivity is obtained from the four oscillation channels. However, due to the large number of muon 
neutrino (and antineutrino) events (see Tab.~\ref{tab:events}), and the relative deviation in 
Fig.~\ref{4prob3}, the main sensitivity comes from $\nu_\mu$ and $\bar{\nu}_\mu$ events and 
some reduced sensitivity from $\bar{\nu}_e$ events. To fully exploit the high energy relative deviations 
that appears in the $\nu_e$ and $\bar{\nu}_e$ events in Fig.~\ref{4prob3}, a high energy flux for DUNE will be considered in the sensitivity analysis of Section~\ref{sec:g32}, and as will be shown, this will substantially improve the sensitivity for testing decoherence.

\begin{table}[!htb]	

	\begin{tabular*}{\columnwidth}{@{\extracolsep{\fill}} c c c c} \hline \hline
		  $\nu_e$-app &  $\bar{\nu}_e$-app &  $\nu_\mu$-disapp &  $\bar{\nu}_\mu$-disapp \\ 
		\hline 
		 1777.69 &  406.025 &  8206.77 &  4124.51 \\ 	\hline \hline
	\end{tabular*}

	\caption{\label{tab:events} Total number of events (signal plus background) for each oscillation 
		channel.}
\end{table}

\subsection{DUNE sensitivity to the decoherence parameters with the default flux configuration} \label{sec:deflim}

In this section we present a sensitivity analysis considering the default flux configuration from \cite{ancillary}. From the previous sections we could see that with the default flux DUNE have a good sensitivity to the decoherence parameters $\Gamma_{21}$ and $\Gamma_{32}$ in a parameter range which is not yet constrained by other experiments. Later, we present a second analysis considering a higher energy flux, which will bring a better sensitivity to $\Gamma_{32}$, since it is the parameter which generates the new peak at $\sim 10$~GeV for the $\nu_e$ appearance probability.

For the analysis presented in this section we have assumed standard oscillation `data' without 
decoherence using values in Table~\ref{eq:parameters-val} and tested the decoherence hypothesis. The 
usual $\chi^2$ analysis have been performed marginalizing over the standard oscillation parameters (except the solar parameters that are kept fixed) 
adding penalties to the $\chi^2$-function with the following standard deviations: $\sigma(\sin^2{(2 
	\theta_{13})})=0.0033$, $\sigma(\sin^2{(2 \theta_{23})})/\sin^2{(2 \theta_{23})}=3\%$, and 
$\sigma(\Delta m^2_{31})/\Delta m^2_{31}=3\%$. The $\delta$ parameter have been also minimized over. 

\begin{figure}[!htb]
	\centering
	\includegraphics[width=0.5\textwidth]{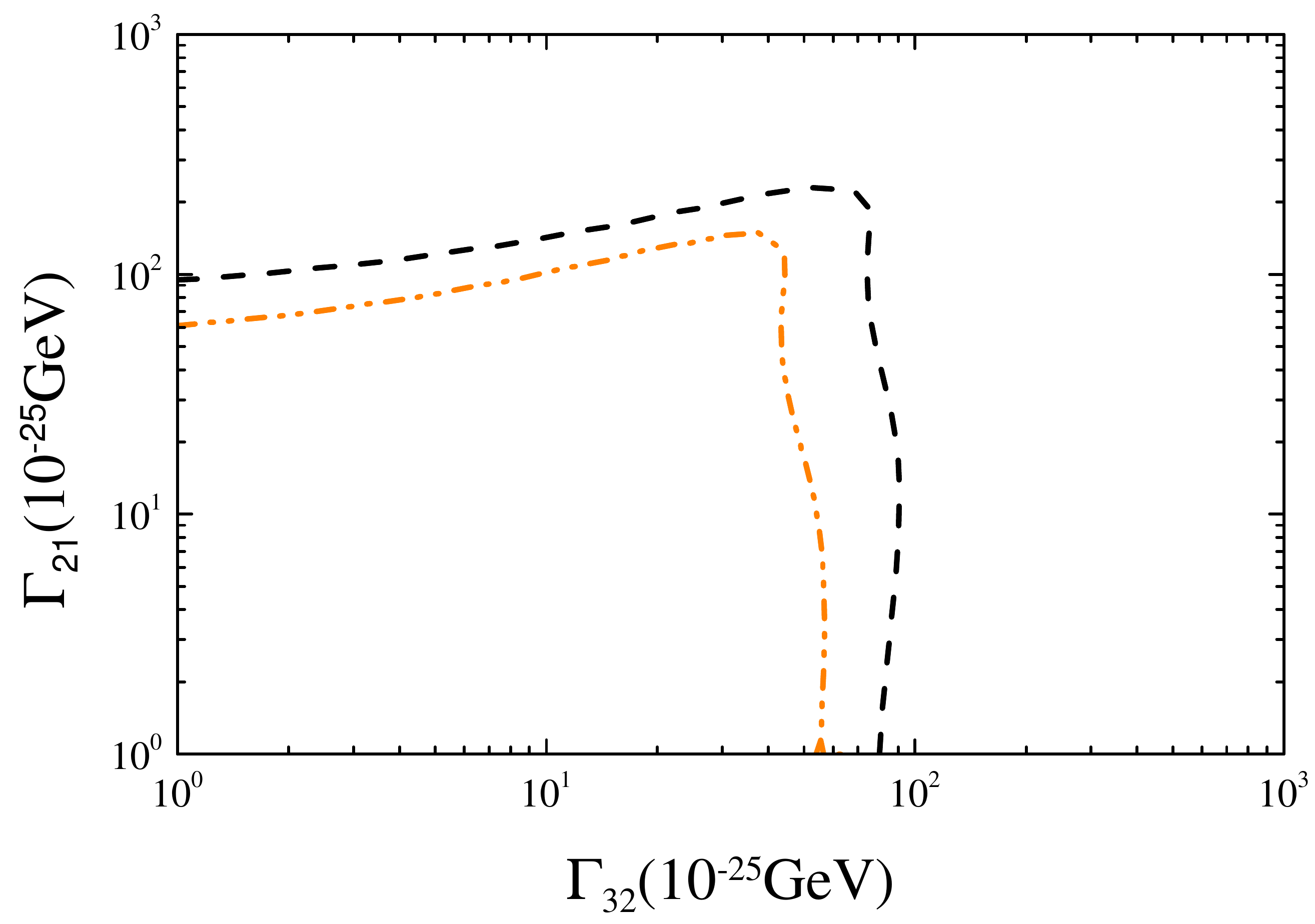}
	\caption{Confidence Level curves at $90\%$ C.L and $3\sigma$ C.L for $2$~d.of.. for the two decoherence parameters $\Gamma_{21}$ and $\Gamma_{32}$, considering the Default Flux given by \cite{ancillary}.}
	\label{conf-1-1}
\end{figure}

Because $\Gamma_{21}, \Gamma_{31}, \Gamma_{32}$ are not all independent, to perform the $\chi^2$ analysis we assumed two of the three decoherence parameters as independent, and defined the other one as a dependent parameter, according to Eqs.~(\ref{g21eq}), Eq.~(\ref{g31eq}) and Eq.~(\ref{g32eq}), making then confidence level curves shown in Fig.~\ref{conf-1-1}.

\begin{figure}[!htb]
	\centering
	\subfloat[$\Delta \chi^2$ \textit{versus} $\Gamma_{32}$, minimizing over $\Gamma_{21}$]{{\includegraphics[height=0.4\textwidth]{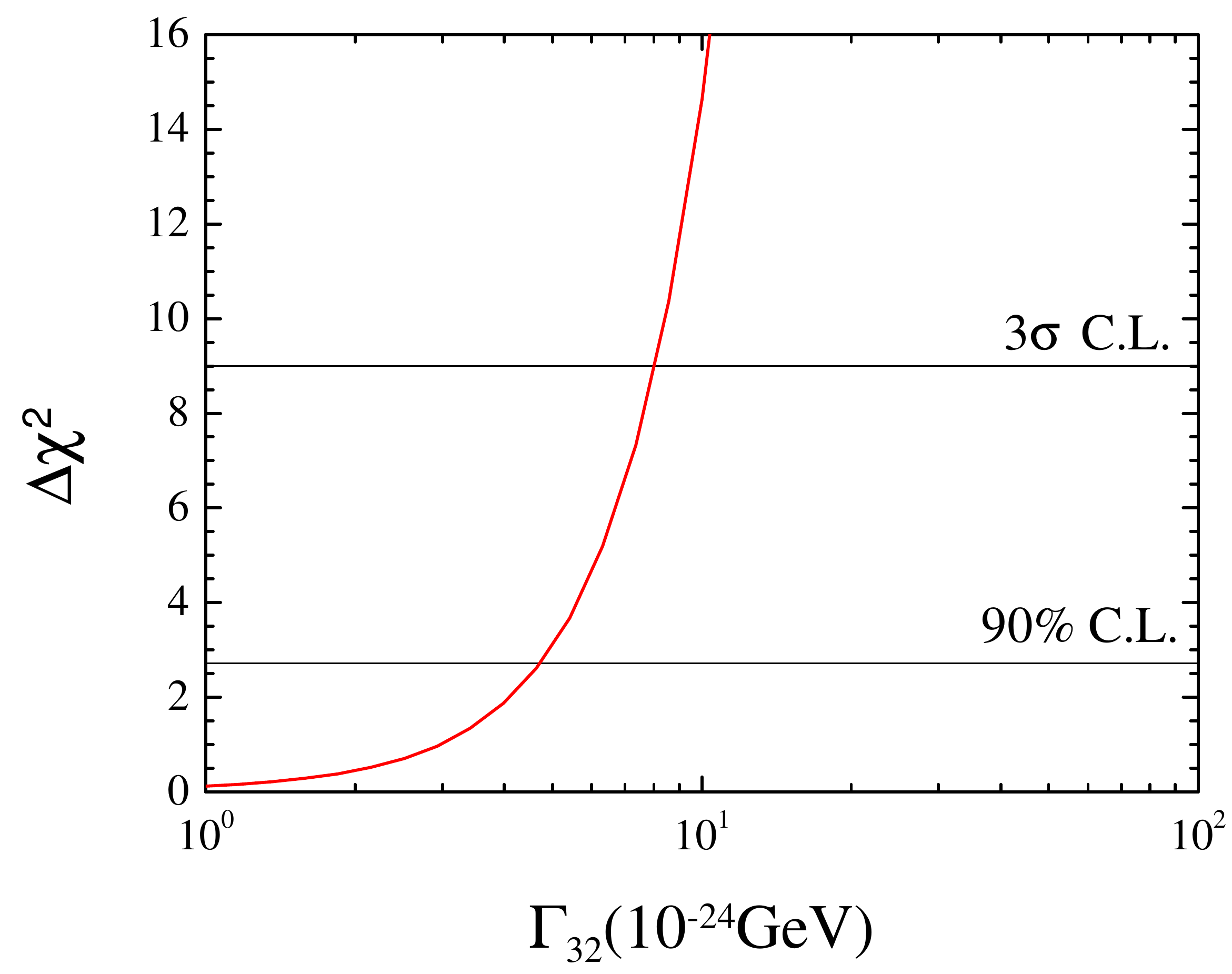}}}%
	\qquad
	\subfloat[$\Delta \chi^2$ \textit{versus} $\Gamma_{21}$, minimizing over $\Gamma_{32}$]{{\includegraphics[height=0.4\textwidth]{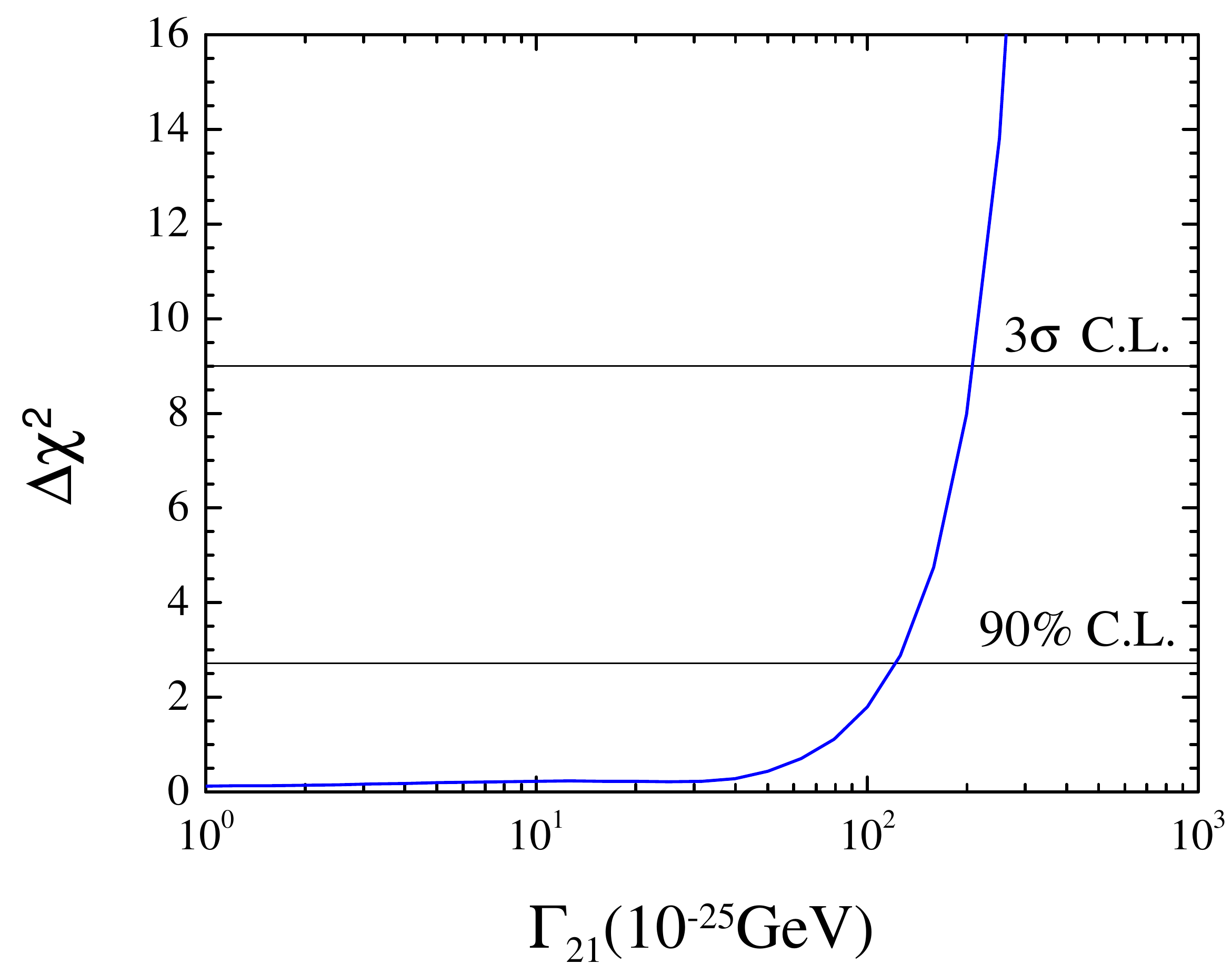}}}%
	\caption{Sensitivity analysis for the Default Flux given by \cite{ancillary}. The horizontal lines define $90\%$ C.L. and $3\sigma$ C.L.}
	\label{proj1-def}
\end{figure}

To obtain the sensitivity regions on each individual parameter we perform the minimization over each of the two decoherence parameters as shown in Fig.~\ref{proj1-def} (a) for $\Delta \chi^2$ \textit{versus} $\Gamma_{32}$ and (b) for $\Delta \chi^2$ \textit{versus} $\Gamma_{21}$. From the $\Delta \chi^2$ profiles we obtained sensitivity regions compiled in table~\ref{tab:constraints}. From the results in table~\ref{tab:constraints} we can see that DUNE has the potential to provide a more stringent limit to $\Gamma_{21}$ than the one given by KamLAND in Ref.~\cite{balikamland}, where the limit for $\Gamma_{21}$ in $95\%$ C.L. is $6.8 \times 10^{-22}$~GeV.

%
%

\begin{table}[!htb]
	\centering
	\begin{tabular*}{\columnwidth}{@{\extracolsep{\fill}}  c  c  c} \hline \hline
		Parameter  & $90\%$ C.L &    $3 \sigma$ C.L \\ 
		\hline
		$\Gamma_{21}\le$ & $1.2\times10^{-23}\,\text{GeV}$ & $2.1\times10^{-23}\,\text{GeV}$\\		
		$\Gamma_{32}\le$ & $4.7\times10^{-24}\,\text{GeV}$& $8.0\times10^{-24}\,\text{GeV}$\\ \hline \hline
	\end{tabular*}
	\caption{\label{tab:constraints} Sensitivity regions for the decoherence parameters from the $\chi^2$ analysis considering the default flux configuration \cite{ancillary}, as shown in Fig.~\ref{proj1-def} (a) and \ref{proj1-def} (b) for 1 d.o.f.}
\end{table}

In the following section we discuss how a high energy neutrino flux, different from the one given by Ref.~\cite{ancillary}, can considerably improve the sensitivity to $\Gamma_{32}$ providing a more suitable configuration to test decoherence at DUNE.

\subsection{Sensitivity analysis for $\Gamma_{32}$ with a high energy flux configuration}\label{sec:g32}

From the discussion at the event level in Section \ref{sec:relr} it is clear that in order to be sensitive to the peak around $10$ GeV in the $\nu_{e}$ appearance channel at DUNE, it is necessary to consider a different flux configuration. Having reached this conclusion, we decided to perform a second sensitivity analysis, but this time considering the High Energy (HE) neutrino flux proposed in Ref.~\cite{bishai}.

\begin{figure}[!htb]
	\centering
	\includegraphics[height=0.35\textwidth]{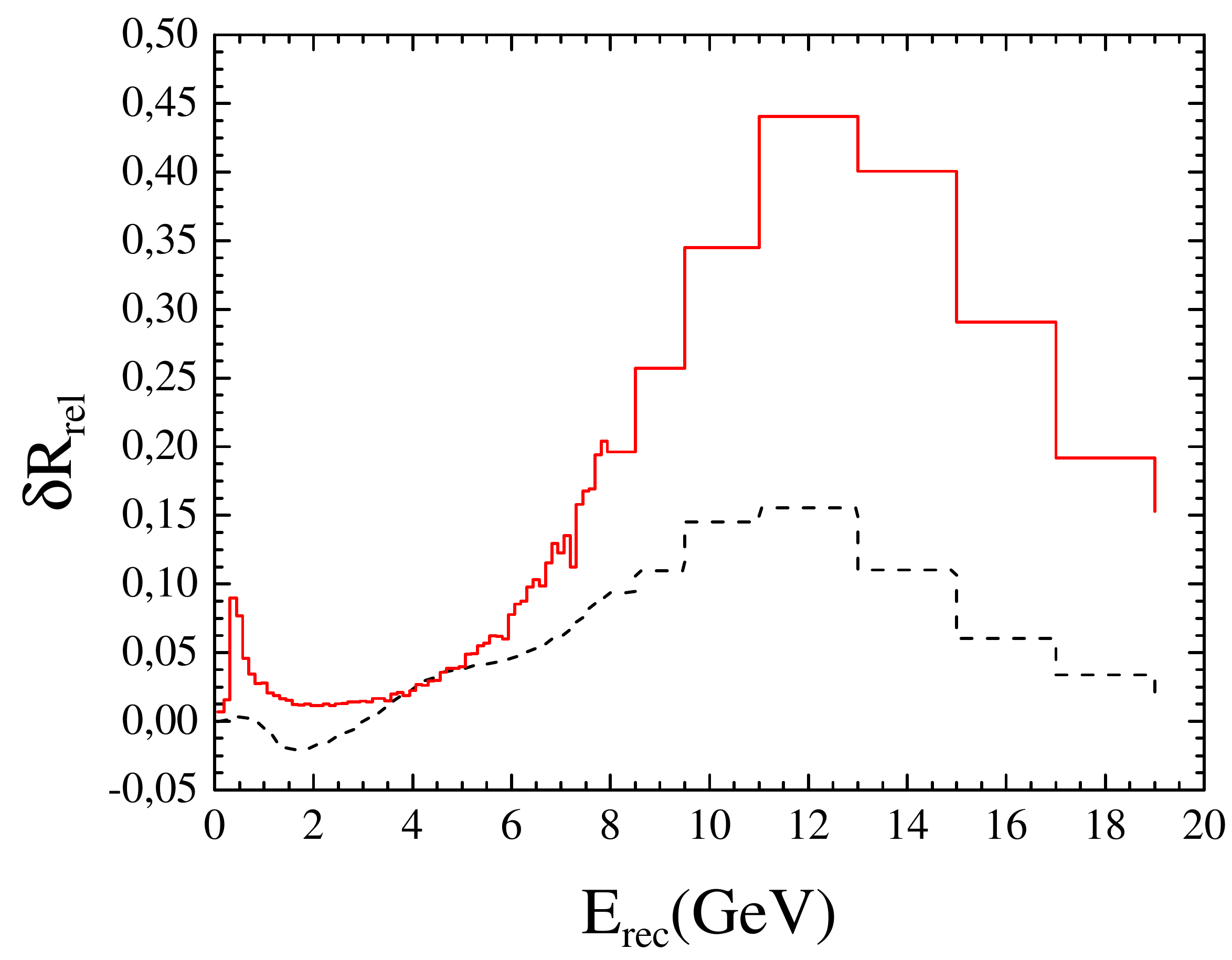}
	
	\caption{Relative deviations respect to the $\nu_e$ appearance events without decoherence ($\Gamma_{32}=0$) for the HE flux from \cite{bishai}. For the event rates with decoherence we considered $\Gamma_{21} = 5.1 \times 10^{-25}$ GeV, $\Gamma_{31} = 3.0 \times 10^{-25}$ GeV, $\Gamma_{32} = 1.6 \times 10^{-24}$~GeV. We also present again the Relative deviations for the $\nu_e$ appearance events considering the default flux \cite{ancillary} for comparison.}
	\label{fig:ev2}
\end{figure}
\begin{figure}[!htb]
	\centering
	\includegraphics[height=0.35\textwidth]{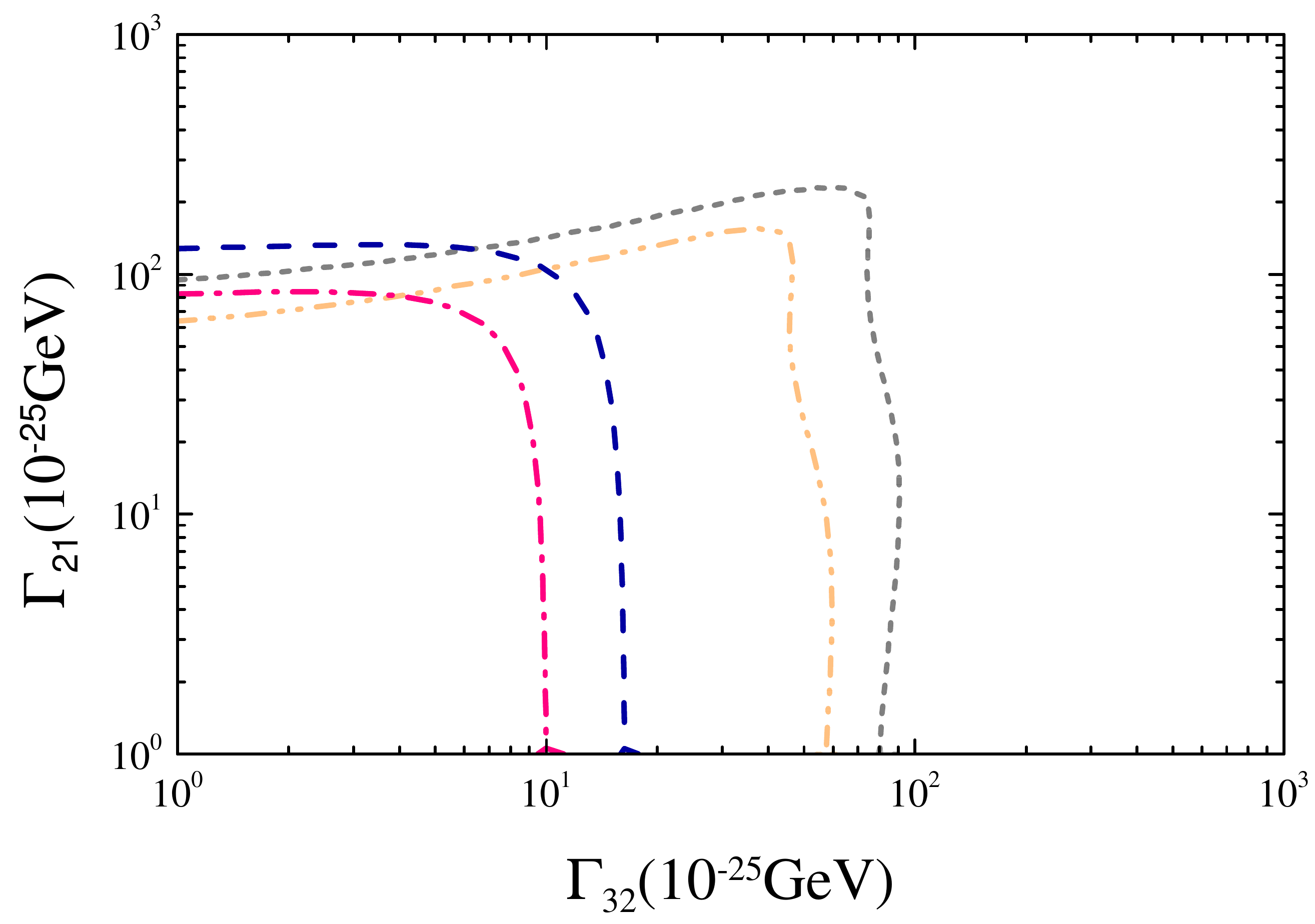}
	\caption{$\Gamma_{32}$ \textit{versus} $\Gamma_{21}$ Confidence Level curves for $90\%$ C.L and $3\sigma$ C.L. considering the HE Flux given by~\cite{bishai}. The Confidence Level curves for the Default Flux \cite{ancillary} are also shown for comparison.}
	\label{conf1-bishai-origin}
\end{figure}
\begin{figure}[!htb]
	\centering
	\includegraphics[height=0.4\textwidth]{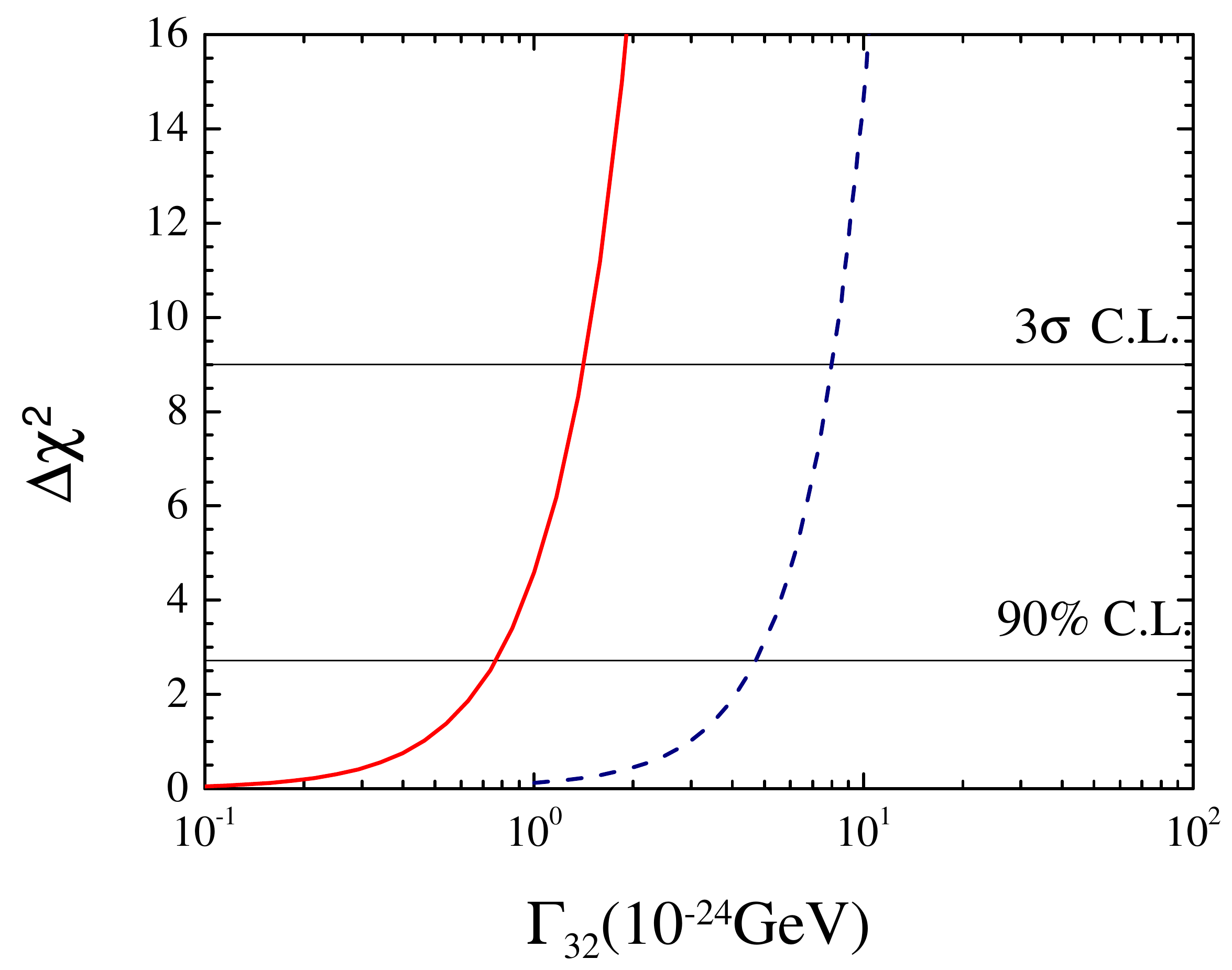}
	\caption{$\Delta \chi^2$ \textit{versus} $\Gamma_{32}$, minimizing over $\Gamma_{21}$, comparing the results between the Default Flux \cite{ancillary} and the HE flux~\cite{bishai}. The horizontal lines define $90\%$ C.L. and $3\sigma$ C.L.}
	\label{proj1-bishai}
\end{figure}

For the sensitivity analysis using the HE flux we excluded the beam contamination from $\nu_e$ and $\overline{\nu}_e$, since we do not have access to this information. Then, we repeated the same procedure of the previous sections, first presenting in Fig.~\ref{fig:ev2} the relative deviation of the number of 
$\nu_e$ events respect to the standard oscillation case without decoherence and finally the sensitivity results. As already 
expected, with the HE flux configuration the peak at in the $\nu_e$ rises to $\sim 45\%$, which suggests that this channel with 
such flux configuration can bring increased sensitivity to the $\Gamma_{32}$ parameter. We showed in 
Section~\ref{sec:probability}, that is the $\Gamma_{32}$ parameter that mostly generates this new peak. Following the same 
procedure of the previous section we performed another sensitivity analysis, and the results are given in 
Fig.~\ref{conf1-bishai-origin} and Fig.~\ref{proj1-bishai}.

From Fig.~\ref{proj1-bishai} we obtained the sensitivity regions compiled in 
Table~\ref{tab:constraints2}, where we present only the limits for $\Gamma_{32}$, since the analysis presented in section~\ref{sec:deflim} already brings the best sensitivity to $\Gamma_{21}$. Sensitivity to $\Gamma_{32}$ given in Table~\ref{tab:constraints2} is enhanced respect to the one found in Section~\ref{sec:deflim}, since the HE flux from~\cite{bishai} is much more suitable to pin down the new peak at $E \sim 10.8$ GeV in the $\nu_{e}$ appearance probability than the default flux~\cite{ancillary}. DUNE has the potential to put a stringent limit to the decoherence parameters, which means that the peak would become much less noticeable in a $\nu_{e}$ appearance probability plot as long as the DUNE measured rates are more `compatible' with standard oscillations.

\begin{table}[!htb]
	\centering
	\begin{tabular*}{\columnwidth}{@{\extracolsep{\fill}}  c  c  c} \hline \hline
		Parameter &  $90\%$ C.L & $3\sigma$ C.L \\ 
		\hline	
		$\Gamma_{32}\le$ & $ 7.7 \times 10^{-25}\,\text{GeV}$ & $1.4\times10^{-24}\,\text{GeV}$\\ \hline \hline
	\end{tabular*}
	\caption{\label{tab:constraints2} Sensitivities to the decoherence parameters from the $\chi^2$ analysis considering the HE flux configuration~\cite{bishai} shown in Fig.~\ref{proj1-bishai} for 1 d.o.f.}
	
\end{table}

\section{Conclusion}

In this work we found sensitivity regions for the decoherence parameters that affect neutrino oscillations in three families considering two possible flux configurations for DUNE.

In Section~\ref{sec:probability} we showed how the new peak at the $\nu_{e}$ appearance probability can be seen as an elimination of a destructive interference, generating then an increase in the transition to $\nu_{e}$. In Section~\ref{results} we showed how the decoherence parameters can be better analyzed by considering different oscillation channels and also different flux configurations, the default flux from~\cite{ancillary} and the HE flux from~\cite{bishai}.

In Section~\ref{sec:deflim} we presented the results for the sensitivity analysis using the flux configuration from Ref.~\cite{ancillary}. In $90\%$ C. L. the sensitivity limits for the parameters given in Table~\ref{tab:constraints} are: $\Gamma_{21}\le 1.2\times10^{-23}\,\text{GeV}$ and $\Gamma_{32}\le 4.7 \times10^{-24}\,\text{GeV}$, and for $3 \sigma$ C.L. the limits are: $\Gamma_{21}\le 2.1\times10^{-23}\,\text{GeV}$and $\Gamma_{32}\le 8.0 \times10^{-24}\,\text{GeV}$. 

As we can see, the limits on $\Gamma_{21}$ are potentially more stringent at DUNE, when compared with the KamLAND experiment, by two orders of magnitude.
On the other hand, DUNE in its default configuration has a reduced sensitivity for $\bar{\nu}_{e}$, suggesting that the limit for $\Gamma_{21}$ comes in most part from the $\nu_{e}$ channel. Therefore, one might think that $\Gamma_{21}$ for $\nu_{e}$ and $\bar{\nu}_{e}$ has some chance to be different. This is the exact scenario for a CPT-like violation such as was proposed in Ref~\cite{gbarenboim} and a new investigation regarding such issue will be presented somewhere else.

Finally, in Section~\ref{sec:g32} we showed how changing to a HE flux configuration DUNE can significantly improve the sensitivity to the $\Gamma_{32}$ parameter potentially pinning down the peak which is the most compelling feature of decoherence at DUNE. The sensitivity regions for such analysis (presented in Table~\ref{tab:constraints2}) are, for $90\%$ C.L.: $\Gamma_{32}\le 7.7\times10^{-25}\,\text{GeV}$, and for $3 \sigma$ C.L. $\Gamma_{32}\le 1.4\times10^{-24}\,\text{GeV}$.

\acknowledgments

The authors would like to thank FAPESP, CAPES and CNPq for several financial supports. 
MMG and PCH are grateful for the support of FAPESP funding Grant
2014/19164-6 and CNPq research fellowships 304001/2017-1 and 310952/2018-2, respectively.
DVF is thankful for the support of the S\~ao Paulo Research Foundation (FAPESP) funding Grants No. 2014/19164-6, 2017/01749-6 and 
2018/19365-2.

\appendix

\section{Conditions on $D_{\mu\nu}$}\label{appendixA}
The Lindblad equation that describes the open system's dynamics is given by:
\begin{equation}
\frac{d\rho(t)}{dt}=-i[H,\rho(t)]
+\frac{1}{2}\sum_{j=1}^{8}\left(\left[V_j,\rho(t) V_j^\dagger\right]+
\left[V_j\rho(t), V_j^\dagger\right]\right)
\end{equation}
where $V_j$ are $3\times 3$ matrices that carry out the new dynamics.
Expanding in Gell-Mann matrices in mass eigenbasis, where $H$ is diagonal, we have:
\[
H=h_3\lambda_3+h_8\lambda_8~~~;~~~ V_j=\sum_{i=1}^8v_{ji}\lambda_i~~~;~~~
\rho=\sum_{i=1}^8\rho_i\lambda_i
\]
where, since $V_j$ is hermitian, all coeficients are real.
The energy conservation condition, $[V_j,H]=0$, leads to:
\begin{eqnarray*}
	[V_j,H]&=&\sum_i\left(h_3v_{ji}[\lambda_i,\lambda_3]+h_8v_{ji}[\lambda_i,\lambda_8]\right)\\
	&=&2i\sum_{i,k}v_{ji}\left(h_3f_{i3k}+h_8f_{i8k}\right)\lambda_k=0
\end{eqnarray*}
The only way to accomplish this with no dependence on $h_{3,8}$ is that
all $v_{ji}$ vanishes except for $v_{j3}$ and $v_{j8}$. Then:
\[
V_j=v_{j3}\lambda_3+v_{j8}\lambda_8
\]
By replacing in the Lindblad equation we obtain:
\begin{widetext}
\begin{eqnarray*}
	\frac{d\rho_k(t)}{dt}\lambda_k&=&2h_m\rho_nf_{mnk}\lambda_k +\frac{1}{2}
	\sum_{j=1}^8\sum_{k=1}^8\sum_{l=3,8}\sum_{m=3,8}
	\rho_kv_{jl}v_{jm}^*\left([\lambda_l,\lambda_k\lambda_m^\dagger]+
	[\lambda_l\lambda_k,\lambda_m^\dagger]\right)\\
	&=&2h_m\rho_nf_{mnk}\lambda_k +\frac{1}{2}\left(\sum_{j}|v_{j3}|^2\right)
	\sum_k\rho_k\left(
	\left[\lambda_3,\lambda_k \lambda_3^\dagger\right]+
	\left[\lambda_3\lambda_k, \lambda_3^\dagger\right]\right) \\
	&+&\frac{1}{2}\left(\sum_{j}|v_{j8}|^2\right)
	\sum_k\rho_k\left(
	\left[\lambda_8,\lambda_k \lambda_8^\dagger\right]+
	\left[\lambda_8\lambda_k, \lambda_8^\dagger\right]\right)\\
	&+&\frac{1}{2}\left(\sum_{j}v_{j3}v_{j8}^*\right)
	\sum_k\rho_k\left(
	\left[\lambda_3,\lambda_k \lambda_8^\dagger\right]+
	\left[\lambda_3\lambda_k, \lambda_8^\dagger\right]\right)\\
	&+&\frac{1}{2}\left(\sum_{j}v_{j3}^*v_{j8}\right)
	\sum_k\rho_k\left(
	\left[\lambda_8,\lambda_k \lambda_3^\dagger\right]+
	\left[\lambda_8\lambda_k, \lambda_3^\dagger\right]\right)
\end{eqnarray*} \end{widetext}

Performing the last sum by direct inspection we get:
\begin{widetext}
\begin{eqnarray*}
	I_1 &=& [\lambda_3,\lambda_k\lambda_3]+[\lambda_3\lambda_k,\lambda_3]\\
	&=&  \lambda_3[\lambda_k,\lambda_3]+[\lambda_3,\lambda_k]\lambda_3\\
	&=&2i(\lambda_3\sum_lf_{k3l}\lambda_l+\sum_lf_{3kl}\lambda_l\lambda_3)\\
	&=&2i\sum_lf_{3kl}[\lambda_l,\lambda_3]=
	-4\sum_{lm}f_{3kl}f_{l3m}\lambda_m=-4\lambda_k\sum_l(f_{3kl})^2\\
	&&\sum_k\rho_kI_1=-(4,4,0,1,1,1,1,0).(\rho_k\lambda_k)
\end{eqnarray*}\end{widetext}
and with a similar procedure:
\begin{eqnarray*}
	I_2&=&  [\lambda_8,\lambda_k\lambda_8]+[\lambda_8\lambda_k,\lambda_8]=
	-4\lambda_k\sum_l(f_{8kl})^2\\
	&&\sum_k\rho_kI_k=-3(0,0,0,1,1,1,1,0).(\rho_k\lambda_k)\\
\end{eqnarray*}
The last two lines can be simplified since $v_{ij}$ are real numbers. In
this case the last two lines can be summed up:
\begin{eqnarray*}
	I_3&=&{\rm [}\lambda_3,\lambda_k\lambda_8{\rm ]}+{\rm [}\lambda_3\lambda_k,\lambda_8{\rm ]}+(3\leftrightarrow 8)=-8\lambda_k\sum_lf_{8kl}f_{3kl}\\
	&&\sum_k\rho_kI_3=-2\sqrt{3}(0,0,0,1,1,-1,-1,0).(\lambda_k\rho_k)
\end{eqnarray*}

Defining the 8-dimensional vectors $\vec{a}_k$ formed by the components $v_{jk}$ in the following way:
\[
(\vec{a}_3)_j\equiv v_{j3}~~~;~~~(\vec{a}_8)_j\equiv \sqrt{3}v_{j8}
\]
we have:
\[
\sum_j|v_{j3}|^2=a_3^2 ~~~;~~~ \sum_j|v_{j8}|^2=\frac{a_8^2}{3} ~~~;~~~
\sum_j(v_{j3}v_{j8}^*)=
\frac{1}{\sqrt{3}}(\vec{a}_3^{\,\dagger}.\vec{a}_8)
\]
we can finally write:
\begin{eqnarray*}
	\frac{d\rho(t)}{dt}&=&-i[H,\rho(t)]+\frac{1}{2}\left(a_3^2I_1+
	a_8^2\frac{I_2}{3}+2\vec{a}_3^{\,\dagger}.\vec{a}_8\frac{I_3}{2\sqrt{3}}\right)
\end{eqnarray*}
and
\begin{widetext}
\begin{eqnarray*}
	D_{mn}  =  -  \frac{1}{2}{\rm diag}(4a_3^2,4a_3^2,0,(\vec{a}_3+\vec{a}_8)^2,
	(\vec{a}_3+\vec{a}_8)^2,(\vec{a}_3-\vec{a}_8)^2,(\vec{a}_3-\vec{a}_8)^2,0)
\end{eqnarray*} \end{widetext}
or
\[
D_{mn}=-{\rm diag}(\Gamma_{21},\Gamma_{21},0,\Gamma_{31},\Gamma_{31},\Gamma_{32},\Gamma_{32},0)
\]
where
\[
\Gamma_{21}\equiv 2a_3^2~~~;~~~\Gamma_{31}\equiv \frac{1}{2}(\vec{a}_3+\vec{a}_8)^2~~~;~~~
\Gamma_{32}\equiv \frac{1}{2}(\vec{a}_3-\vec{a}_8)^2
\]

For simplicity we treated $\vec{a}_3$ and $\vec{a}_8$ as colinear, so they were
treated as scalars. So the conclusion is that the matrix energy conservation
in the neutrino sector requires a diagonal format for $D_{mn}$, with a specific
relation between its terms.

\section{Decoherence in matter and positivity}\label{appendix}

The authors of Ref.~\cite{Carpio:2017nui} claim that decoherence cannot be defined in the effective mass basis, and 
that (apart from very specific cases) the forms of the dissipative matrices in vacuum and in matter cannot be the same. 
In this appendix we comment that, under certain conditions, decoherence can be defined as arising from the same matrices 
in both contexts, such that it preserves a physical interpretation where the decoherence effect acts only on the quantum 
interference terms, such as was discussed in this paper and also in Refs.~\cite{balikamland,robertosun}.

The dissipator in Eq.~(\ref{dissipdune}) is obtained when it is imposed that
\begin{equation}
[H_S,V_{k}]=0,
\label{cond1app}
\end{equation} where $H_S$ is the Hamiltonian of the subsystem.

Since the subsystem is different when one considers neutrinos in vacuum or in matter, if $V_k$ has the same form in both 
bases, then Eq.~\ref{cond1app} is not satisfied at the same time in such cases. This is exactly what happens in 
Refs.~\cite{lisi, Carpio:2017nui}. As it is shown in \cite{robertosun}, this implies that the decoherence effect and the 
so called relaxation effect cannot be fully separated. In fact, as argued in Refs.~\cite{balikamland, robertosun}, the 
work in~\cite{lisi} finds constraints for the relaxation effect, or for decoherence in a model dependent approach. It is 
important to point out that decoherence is an effect which acts only on the quantum interference terms of the 
oscillation probabilities, while relaxation acts only on the constant terms, which allow flavor conversion even without 
mixing between the neutrino families.

For two neutrino families~\cite{robertosun} we have that the parameterization:
\begin{equation}
\tilde{V}_{k}=\sqrt{\gamma_{1}}
\left(\begin{array}{cc}
\cos\delta_\theta & -\sin\delta_\theta \\ 
-\sin\delta_\theta & -\cos\delta_\theta
\end{array}\right),
\label{robapp1}
\end{equation} 
with $\delta_\theta = 2(\tilde\theta-\theta)$, such that $\tilde\theta$ is the effective mixing angle in matter, leaves eq.(\ref{cond1app}) unchanged for any matter density.

As we can see, in vacuum Eq.~(\ref{robapp1}) assumes the following form:

\begin{equation}
V_{k}=\sqrt{\gamma_{1}}
\left(\begin{array}{cc}
1 & 0 \\ 
0 & -1
\end{array}\right),
\label{robapp2}
\end{equation} since in vacuum $\delta_\theta = 0$.

To assure that the condition Eq.~(\ref{cond1app}) is satisfied for neutrinos propagating in both vacuum and in constant density matter, it is shown in Ref.~\cite{robertosun} that the operators $V_k$ must also transform when there is a change of basis. Therefore we must have that:
\begin{equation}
\tilde{V}_{k}=U_T^\dagger V_{k} U_T = \sqrt{\gamma_{1}}
\left(\begin{array}{cc}
1 & 0 \\ 
0 & -1
\end{array}\right),
\label{transfvk}
\end{equation} where $U_T=U^\dagger U_M$, and $U_M$ is the rotation matrix between the flavor basis and the effective mass basis, and as we can see, it is equal to Eq.~(\ref{robapp2}).

When $V_k$ transform as Eq.~(\ref{transfvk}) the dissipator in Eq.~(\ref{dissipdune}) (where we only consider 
decoherence, not relaxation) is valid for neutrinos propagating in both vacuum and in matter. Since the form is the 
same, the conditions for positivity are also the same for both cases, which assures that when Eqs.~(\ref{g21eq}) -- 
(\ref{g32eq}) are obeyed the physical meaning of the probabilities are guaranteed for oscillation both in vacuum and in 
matter. It is also important to point out that, different from what is assumed by Ref.~\cite{Carpio:2017nui}, in this 
work decoherence is assumed to be dependent on the matter density, as can be seen from Eq.~(\ref{robapp1}). More details 
of this discussion can be found in Ref.~\cite{robertosun} for the case of two neutrinos.

It is worth to notice that, even though the calculations presented in Ref.~\cite{robertosun} were made for the case of two neutrinos, the discussion of the concepts involved is very general, and its conclusions can be extended to the three neutrino case.

\bibliography{artigodecdune}

%
%
\end{document}